\documentclass{emulateapj}
\usepackage{amsmath}
\input{epsf}
\usepackage{graphicx}
\usepackage{epstopdf}
\usepackage{bm}
\usepackage[breaklinks,colorlinks, urlcolor=blue,citecolor=blue,linkcolor=blue]{hyperref}
\usepackage{natbib}
\usepackage{verbatim}
\DeclareMathAlphabet{\mathpzc}{OT1}{pzc}{m}{it}

\usepackage{amssymb}
\usepackage{rotating}
\usepackage{gensymb}
\newcommand{\GG}[1]{}
\def\lya{Ly$\alpha$~} 
\def\kms{km s$^{-1}$} 
\def\kpc{kpc}
\def\Mpc{Mpc}
\def\ar{arcmin}
\def\lyb{Ly$\beta$ }
\def\rhobar{\overline{\rho}}
\newcommand{\ben}{\begin{enumerate}}
\newcommand{\een}{\end{enumerate}}

\shorttitle{Quasar environment using proximity effect} 
\shortauthors{Jalan, Chand \& Srianand}

\begin{document}

\title{Probing the environment of high-z quasars \\using the proximity
  effect in projected quasar pairs}

\author{Priyanka Jalan$^{1,2}$,
 Hum Chand$^1$,
  Raghunathan Srianand$^3$}

\affil{$^1$ Aryabhatta Research Institute of Observational Sciences (ARIES),
 Manora Peak, Nainital$-$ 263002, India \\
$^2$ Department of Physics and Astrophysics, University of Delhi, Delhi 110007, India: \url{priyajalan14@gmail.com}\\ 
$^3$  Inter-University Centre for Astronomy and
Astrophysics (IUCAA), Postbag 4, Ganeshkhind, Pune 411 007,
India\\}

\begin{abstract}
 We have used spectra of 181 projected quasar pairs at separations
 $\le 1.5$ \ar~from the Sloan Digital Sky-Survey Data Release 12 in
 the redshift range of 2.5 to 3.5 to probe the proximity regions of
 the foreground quasars. We study the proximity effect both in the
 longitudinal as well as in the transverse directions, by carrying out
 a comparison of the \lya absorption lines originating from the
 vicinity of quasars to those originating from the general
 inter-galactic medium at the same redshift. We found an enhancement
 in the transmitted flux within 4 Mpc to the quasar in the
 longitudinal direction. However, the trend is found to be reversed in
 the transverse direction. In the longitudinal direction, we derived
 an excess overdensity profile showing an excess up to $r\leq 5$ Mpc
 after correcting for the quasar's ionization, taking into account the
 effect of low spectral resolution. This excess overdensity profile
 matches with the average overdensity profile in the transverse
 direction without applying any correction for the effect of the
 quasar's ionization.  Among various possible interpretations, we
 found that the anisotropic obscuration of the quasar's ionization
 seems to be the most probable explanation. This is also supported by
 the fact that all of our foreground quasars happen to be Type-I
 AGNs. Finally, we constrain the average quasar's illumination along
 the transverse direction as compared to that along the longitudinal
 direction to be $\leq$27\% (3$\sigma$ confidence level).
 
\end{abstract}

\keywords{quasars: absorption lines --- objects: general ---
  intergalactic medium --- techniques: spectroscopic --- Methods: data
  analysis - proximity effect - statistical}

\section{Introduction}
\label{s:Intro}
The redshifted H~{\sc i} \lya absorption lines seen in the spectra of
distant quasars (commonly known as \lya forest), are powerful probes
of the physical conditions in the inter-galactic medium \citep[IGM,
][]{Gunn1965ApJ...142.1633G,Sargent1980PhyS...21..753S,Rauch1998ARA&A..36..267R,Romeel2010MNRAS.408.2051D,Pieri2010ApJ...724L..69P}
and the parameters of the background cosmology
\citep[][]{Cocke1989ApJ...346..613C,Weinberg1997ApJ...490..564W,Croft1998ApJ...495...44C,Seljak2003MNRAS.342L..79S,Tytler2004ApJ...617....1T,Fechner2007A&A...463...69F,Busca2013A&A...552A..96B,Delubac2015A&A...574A..59D}.
It is believed that most of the \lya lines with column density,
$N_{\rm HI}\leq 10^{14}$ cm$^{-2}$, originate from quasi-linear
density fluctuations in which the hydrogen gas is in ionization
equilibrium with the meta-galactic UV background (UVB) radiation
produced by star-forming galaxies and quasars
\citep[][]{Bergeron1990A&A...235....8B,Madau2000MNRAS.312L...9M,Meiksin2003MNRAS.342.1205M,Bolton2006MNRAS.366.1378B,Hopwood2010ApJ...716L..45H,Haardt2012ApJ...746..125H,Khaire2015MNRAS.451L..30K,khaire2015ApJ...805...33K}. As
non-linear effects are sub-dominant, properties of the gas responsible
for the \lya forest can be well described using few basic ingredients
such as quasi-linear theory for the growth of baryonic structure,
ionization equilibrium with UVB radiation field, and the effective
equation of state of the gas
\citep[][]{Bi1993ApJ...405..479B,Muecket1996A&A...308...17M,Bi1997ApJ...479..523B,Hui1997astro.ph..2167H,Weinberg1999ApJ...522..563W,Schaye2001astro.ph.12022S,Schaye2001ApJ...559..507S,Rollinde2001A&A...376...28R,Viel2002MNRAS.336..685V,Kim2002MNRAS.335..555K,Lehner2007ApJ...658..680L}. This
basic idea is also confirmed by many hydrodynamical simulations
\citep[][]{Cen1994ApJ...437L...9C,Hernquist1996ApJ...457L..51H,Wadsley1997ASPC..123..332W,Zhang1997ApJ...485..496Z,Theuns1998MNRAS.301..478T,Machacek2000ApJ...532..118M,Efstathiou2000RSPTA.358.2049E,Choudhury2001MNRAS.322..561C,Choudhury2001ApJ...559...29C,Regan2007MNRAS.374..196R,White2010ApJ...713..383W,Ozbek2016MNRAS.456.3610O,Sorini2016ApJ...827...97S,Bolton2017MNRAS.464..897B}.
Therefore, the \lya forest has been extensively used to derive many
cosmological properties such as matter power spectrum, baryon density
in the universe, temperature and ambient radiation field
\citep[][]{Hui1999ApJ...517..541H,Nusser1999MNRAS.303..179N,Pichon2001MNRAS.326..597P,Croft2002ApJ...580..634C,McDonald2001ApJ...562...52M,Viel2006MNRAS.365..231V,Borde2014JCAP...07..005B,Gaikwad2018MNRAS.474.2233G}
by comparing properties of simulated and observed data. Under the
photoionization equilibrium, the optical depth ($\tau$) of the \lya
absorption in the IGM is related to the overdensity of the gas,
$\Delta\equiv \rho/\langle\rho\rangle$ where, $\langle\rho\rangle$
being the mean IGM density, \citep[e.g., see][and references
  therein]{Rollinde2005MNRAS.361.1015R} as,
\begin{equation}
\tau~\propto~\Delta^2 [T(\Delta)]^{-0.7}/\Gamma_{\textsc{Hi}}\propto
\Delta^{2-0.7(\gamma-1)}/\Gamma_{\textsc{Hi}}\,.
\label{e:eqn1}
\end{equation}
Here, $\Gamma_{\textsc{Hi}}$ is the hydrogen photoionization rate,
$T(\Delta)$ is the temperature of the gas at overdensity $\Delta$,
given by the temperature-density relation, $T=T_0\Delta^{\gamma-1}$,
with an exponent $\gamma$ and $T_0$ being the temperature at mean IGM
density (i.e., $\Delta \sim $1). \par Observationally, an independent
way of estimating H~{\sc i} photoionization rate
($\Gamma_{\textsc{Hi}}$) is by the analysis of the \lya absorption
sufficiently close to the quasar. In this case, the UV-field is
dominated by the radiation originating from the quasars which lead to
a deficit of detectable \lya absorption lines. As a result, contrary
to the expectation that the amount of \lya absorption should be an
{\em increasing} function of redshift, a reversal of trend is seen for
absorption redshifts close to the emission redshift of the
quasar. This effect is commonly known as inverse or the {\em proximity
  effect} \citep[e.g.,
][]{Carswell1982MNRAS.198...91C,Murdoch1986ApJ...309...19M,Tytler1987ApJ...321...69T,Bajtlik1988ApJ...327..570B,Kulkarni1993ApJ...413L..63K,Bechtold1994ApJS...91....1B,srianand1996MNRAS.280..767S,Cooke1997MNRAS.284..552C,Liske2001MNRAS.328..653L,Worseck2006A&A...450..495W,Faucher2008ApJ...673...39F,Wildlpe12008MNRAS.388..227W,Prochaska2013ApJ...776..136P,Khrykin2016ApJ...824..133K}. The
strength of this effect depends on the ratio of the ionization rates
contributed by the quasar and the UVB radiation. Hence, based on the
extent of the proximity region along with the fact that the H~{\sc i}
photoionization rate due to quasar radiation can be determined
directly from its observed luminosity, one can infer the value of
$\Gamma_{\textsc{Hi}}$ of the UVB. This method of using the line of
sight proximity effect (i.e., longitudinal proximity effect) was
pioneered by \citet[][]{Bajtlik1988ApJ...327..570B}. Subsequent
studies have yielded a wide variety of $\Gamma_{\textsc{Hi}}$
estimates varying from 1.5 to 9 in units of $10^{-12}$ s$^{-1}$ at
$z=3$
\citep[][]{Lu1991ApJ...367...19L,Kulkarni1993ApJ...413L..63K,Cristiani1995MNRAS.273.1016C,Giallongo1996ApJ...466...46G,Cooke1997MNRAS.284..552C,Scott2000ApJS..130...67S,Liske2001MNRAS.328..653L,Agliolpe32008AA...480..359D,Calverley2011MNRAS.412.2543C,Partl2011MNRAS.415.3851P,Syphers2013arXiv1310.1616S}.
\par Most of the previous measurements of $\Gamma_{\textsc{Hi}}$ using
the proximity effect have assumed that the distribution of absorbing
gas close to the quasar is same as that in the general IGM. This
assumption may not be valid in a scenario where galaxies as well as
IGM tends to cluster around the quasars
\citep[][]{Bahcall1969ApJ...157L..77B,Hartwick1990ARA&A..28..437H,Bahcall1991ApJ...380L...9B,Fisher1996ApJ...468..469F,srianand1996MNRAS.280..767S,Fukugita2004ApJ...603L..65F,Croom2005MNRAS.356..415C,Rollinde2005MNRAS.361.1015R,Adams2015MNRAS.448.1335A,Eftekharzadeh2017MNRAS.468...77E}. As
a result, previous measurements might have underestimated the
magnitude of the proximity effect, or equivalently, overestimated the
$\Gamma_{\textsc{Hi}}$ \citep[see
  also,][]{Loeb1995ApJ...448...17L,Faucher2008ApJ...673...39F}.
According to the hierarchical models of the galaxy formation, the
super-massive black holes that are thought to power quasars reside in
massive halos
\citep[][]{Miralda1996ApJ...471..582M,Wyithe2006MNRAS.366.1029W,Shen2007AJ....133.2222S,Kim2008MNRAS.387..377K,White2012MNRAS....424.933W,Andreu1475-7516-2013-05-018,Rodriguez2017MNRAS.468..728R},
which are strongly biased towards high-density regions, especially at
the higher redshifts. Therefore, it is expected that the gas in the
neighborhood of the quasars must have a higher density than the gas in
the general IGM at the same epoch (i.e., at same redshift). \par While
most such studies using \lya forest were done in the longitudinal
direction, the environment of a quasar can also be probed in the
transverse direction using quasar pairs, commonly known as transverse
proximity effect
\citep[][]{Adelberger2004ApJ...612..706A,Schirber2004ApJ...610..105S,Rollinde2005MNRAS.361.1015R,Worseck2007A&A...473..805W,Goncalves2008ApJ...676..816G,Gallerani2011JPhCS.280a2008G,Hennawi2013ApJ...766...58H,Schmidt2018ApJ...861..122S}.
The main principle here is to use \lya absorption lines detected along
the sight-line of a background quasar, near the redshift of the
foreground quasar, to probe the ionization effect due to a foreground
quasar in the transverse direction. The absorption seen in the
spectrum of the background quasar at the redshift of the foreground
quasar will be influenced by the excess radiation and overdense
environment in which the foreground quasar resides \citep[e.g.,
  see][]{Hennawi2006ApJ...651...61H,Hennawi2007ApJ...655..735H,Prochaska2009ApJ...690.1558P,Hennawi2013ApJ...766...58H,Prochaska2014ApJ...796..140P,Lau2016ApJS..226...25L,Lau2018ApJ...857..126L}. Many
recent studies have also found a significantly more absorption close
to the quasars than what is expected in the IGM and concluded that the
quasars reside in regions having gas density greater than the typical
gas density of the IGM
\citep[][]{Croft2004ApJ...610..642C,Schirber2004ApJ...610..105S,Rollinde2005MNRAS.361.1015R,Guimaraes2007MNRAS.377..657G,Kirkman2008MNRAS.391.1457K,Finley2014A&A...572A..31F,Adams2015MNRAS.448.1335A}.
Recently, \citet[][]{Prochaska2013ApJ...776..136P} used a sample of
650 projected quasar pairs to study the transverse proximity effect of
luminous quasars (at $z\sim 2$) at proper separations ranging from 30
\kpc~to 1 \Mpc. Based on anisotropic absorption \citep[also see,
][]{Prochaska2014ApJ...796..140P,Lau2016ApJS..226...25L,Lau2018ApJ...857..126L}
found in their analysis around quasars they concluded that the gas in
the transverse direction is likely to be less illuminated by ionizing
radiation compared to that along our line-of-sight to the quasars
\citep[see
  also,][]{Bowen2006ApJ...645L.105B,Farina2013MNRAS.429.1267F}. \par
Another aspect one has to consider while studying the transverse
proximity effect is the effect of a finite lifetime or an episodic
quasar phase. Here, the extra light travel time in the perpendicular
direction between the two quasar sightlines (i.e., $r_\perp$/c) may
lead to a situation where the effect of excess radiation may be weaker
or absent along line-of-sight of the background quasar sightline in
the perpendicular direction from the foreground quasar when we see it
in its initial stages of the active phase.  Given the important
implications of such studies, it is imperative to use a sample of
projected quasar pairs (henceforth quasar pairs) at smaller angular
separation ($<$ 1.5~\ar), to probe the quasars environment even at
\kpc~scales, where both the ionization effect, as well as the
overdensity effect, might be appreciable. However, to lift the
degeneracy due to the ionization and/or the overdensity on the
observed optical depth, the measurement of UVB radiation using an
independent method will be needed. For this purpose, we use the recent
estimate of UVB radiation based on the updated comoving specific
galaxy and quasar emissivity at different frequencies (from UV to FIR)
and redshifts
\citep[][]{Khaire2015MNRAS.451L..30K,Khaire2019MNRAS.484.4174K}. Particularly,
it will help to infer any difference in the optical depth between the
longitudinal and the transverse direction, to confirm or refute the
validity of the assumed isotropic quasar emission and constraint the
excess overdensity. \par On the other hand, the advent of large quasar
surveys such as Sloan Digital Sky-Survey Data Release 12 \citep[SDSS
  DR12, e.g., see][]{Paris2017A&A...597A..79P} has enabled us to
gather a large sample of quasar pairs with a small separation of less
than 1.5-\ar. As a result, we can utilize these low/moderate
resolution spectra of this large set of quasar pairs to probe the
quasar environment and/or anisotropic emission from the quasars. In
particular, it is now possible to carry out analysis using a control
sample of \lya absorption at the same epoch having a spectrum with a
similar signal to noise and spectral resolution to that of the quasar
pairs, which forms the main motivation of this work. \par The paper is
organized as follows. In Sect.~\ref{s:data} we describe our sample and
selection criteria used to make the sample for the study of both the
longitudinal and the transverse proximity effect along with the
details of our control sample. In Sect.~\ref{s:Analysis}, we present
pixel optical depth analysis and results along with detailed
discussions on the validation of appropriate ionization corrections
for the pixel optical depth from low/moderate resolution SDSS spectra
using simulated spectra. The discussions are presented in
Sect.~\ref{s:Discussion}. Finally, the summary and conclusions are
given in Sect.~\ref{s:summary}.  Throughout, we have used a flat
background cosmology with $\Omega_m$ = 0.286, $\Omega_\lambda$ =
0.714, and $H_o$ = 69.6 \kms \Mpc~$^{-1}$
\citep[][]{Bennett2014ApJ...794..135B}. All the distances mentioned in
our paper are proper distances unless noted otherwise.

\begin{figure}	
   \centering
   \includegraphics[height=7cm,width=8cm]{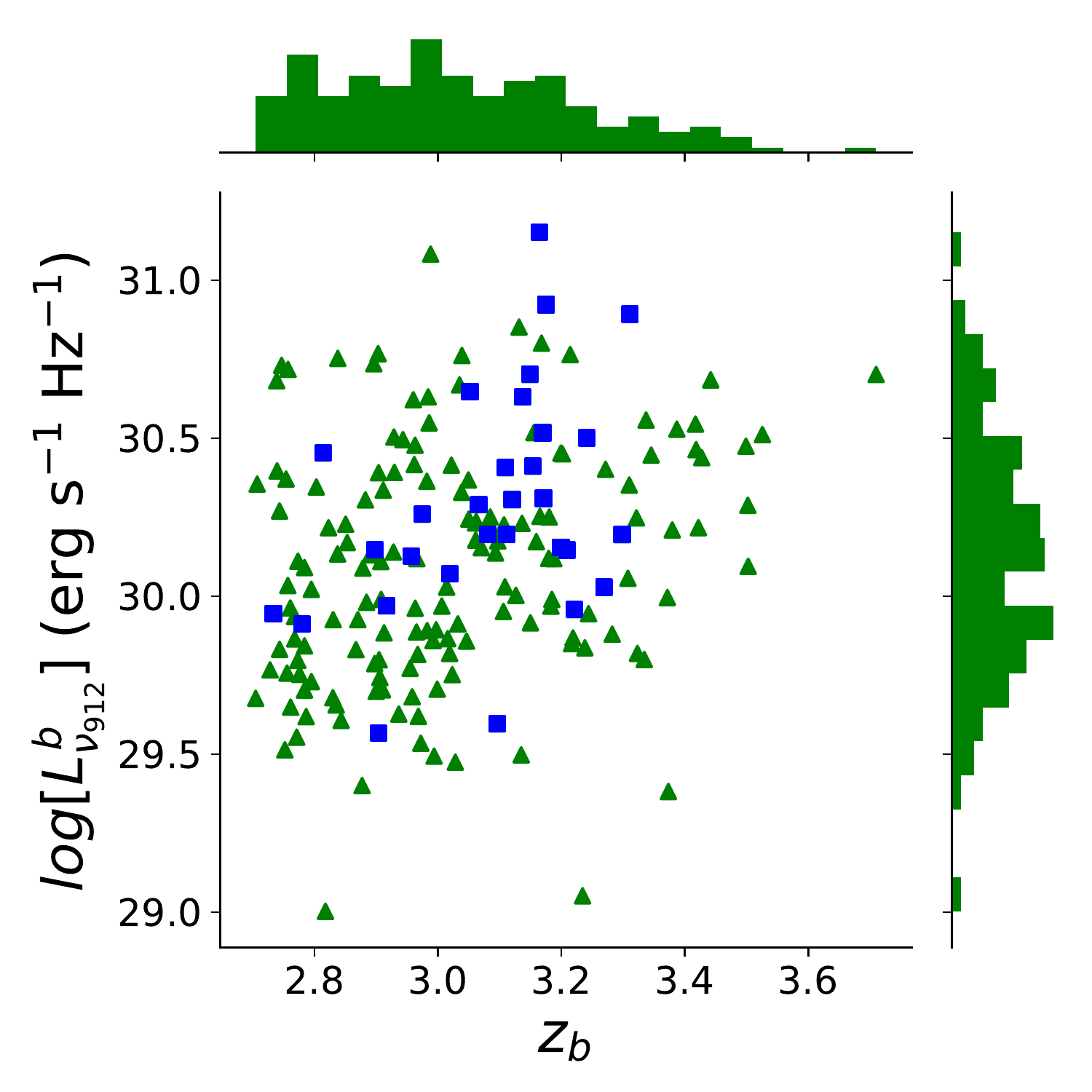}
   \includegraphics[height=7cm,width=8cm]{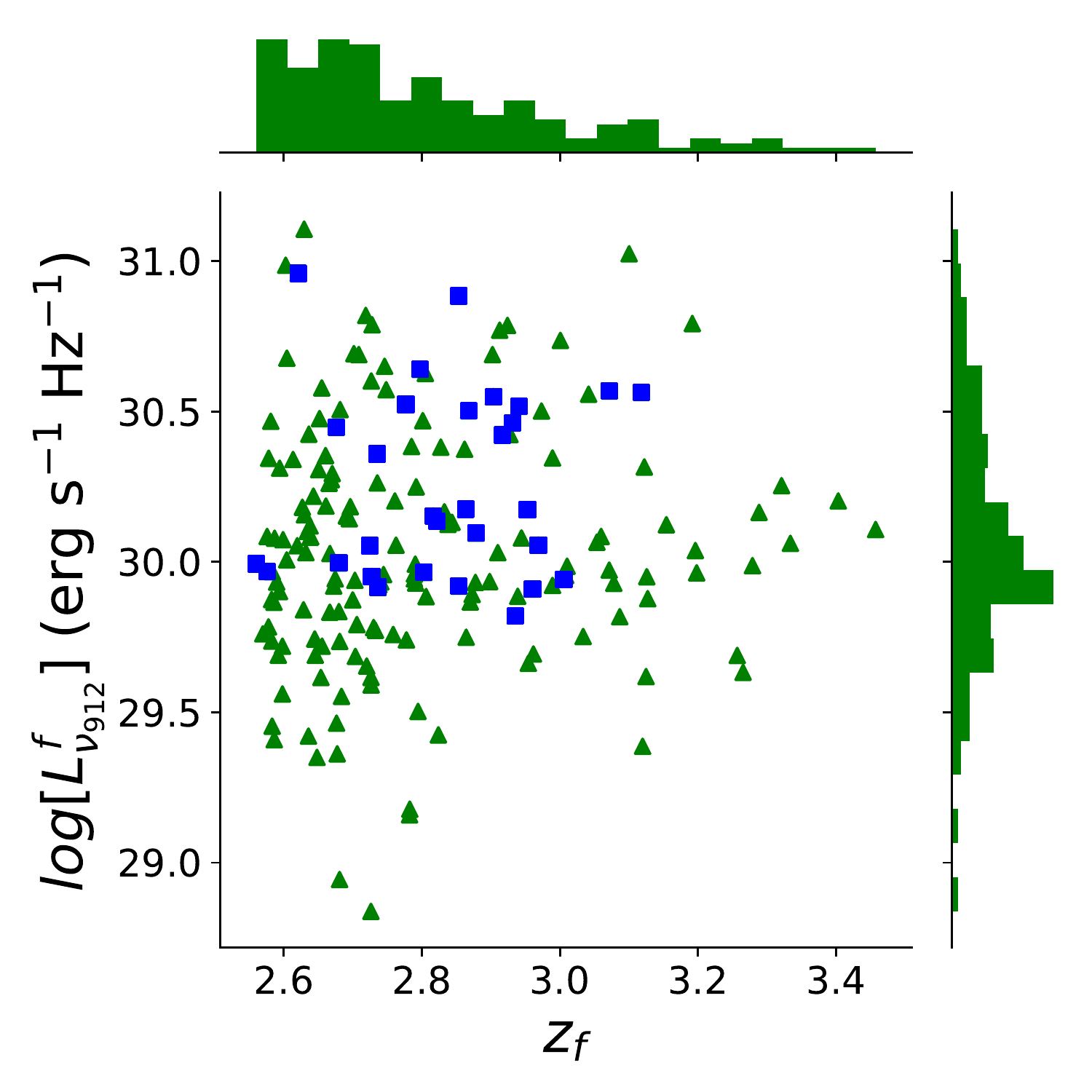}
   \includegraphics[height=7cm,width=8cm]{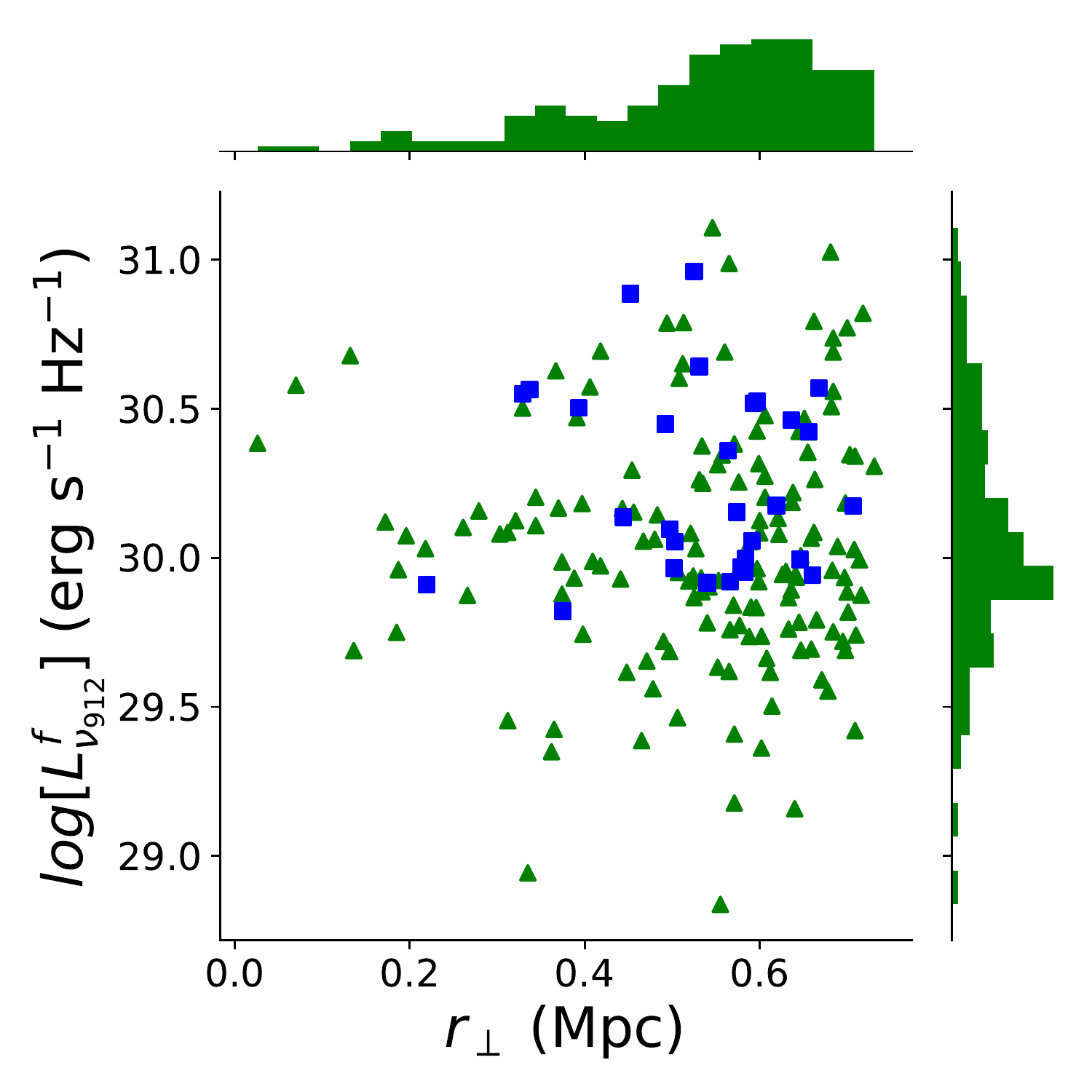}
   \caption {Some properties of the quasars in our sample. {\it Top:}
     Luminosity at 912 \AA~(log[$L_{\nu_{912}}$] in units of
     erg~s$^{-1}$ Hz$^{-1}$) of background quasars (listed in
     Table~\ref{table:sources}) versus $z_{b}$. {\it Middle:} Same as
     top but for the foreground quasars. {\it Bottom:} $L_{\nu_{912}}$
     of the foreground quasars versus the perpendicular distance
     between the foreground and the background quasars at
     $z_{f}$. The top and right side histograms show the
       distribution of parameters labeled at abscissa and ordinate
       respectively. The blue squares are the quasars found in common
     with the sample of \citet[][]{Prochaska2013ApJ...776..136P}.}
      \label{Fig:qso_prop}
\end{figure}
\begin{table*}
  \centering
      {
        \setlength{\tabcolsep}{1.2pt}
\caption{{ Basic properties of 181 quasar pairs in our sample.}
\label{tab:source_info}}
\begin{tabular}{c@{\hspace{0.35cm}}c@{\hspace{0.35cm}}c@{\hspace{0.35cm}}ccc@{\hspace{0.35cm}}c@{\hspace{0.35cm}}c@{\hspace{0.35cm}}c@{\hspace{0.35cm}}ccc} 
\hline
 \multicolumn{1}{c}{SN.}  & \multicolumn{1}{c}{Background quasar}  &
 \multicolumn{1}{c}{$z_{b}$}  & \multicolumn{1}{c}{$\delta z_{b}$} &
 \multicolumn{1}{c}{log($L^{b}_{\nu_{912}})$} &  \multicolumn{1}{c}{$m^b_r$} &
 \multicolumn{1}{c}{Foreground quasar}  &
 \multicolumn{1}{c}{$z_{f}$} &    \multicolumn{1}{c}{$\delta z_{f}$} &
 \multicolumn{1}{c}{log($L^{f}_{\nu_{912}})$} &  \multicolumn{1}{c}{$m^f_r$} & \multicolumn{1}{c} {$r_{\perp}$ }
 \\
\multicolumn{1}{c}{}  & \multicolumn{1}{c}{}  & \multicolumn{1}{c}{}  &
\multicolumn{1}{c}{}  & \multicolumn{1}{c}{(erg~s$^{-1}$ Hz$^{-1}$)} & \multicolumn{1}{c}{} &
  \multicolumn{1}{c}{}  & \multicolumn{1}{c}{} & \multicolumn{1}{c}{} &
 \multicolumn{1}{c}{(erg~s$^{-1}$ Hz$^{-1}$)} & \multicolumn{1}{c}{}  &  \multicolumn{1}{c} {(\kpc)}
 \\
 \\
 \hline
   1  &   J000244.88+125757.60  & 3.334   &   0.005    & 29.800  & 21.046  & J000243.44+125830.00   &   3.154  &   0.005   &  30.124  &  20.998   & 321 \\
   2  &   J001119.92+025840.80  & 3.503   &   0.005    & 30.095  & 21.120  & J001123.28+025826.40   &   3.279  &   0.005   &  29.988  &  21.154   & 409 \\
   3  &   J001431.20+073224.00  & 2.882   &   0.004    & 30.305  & 20.940  & J001431.92+073332.40   &   2.594  &   0.004   &  30.312  &  20.133   & 552 \\
   4  &   J001609.12+103126.40  & 2.955   &   0.004    & 29.772  & 21.657  & J001606.48+103155.20   &   2.824  &   0.004   &  29.425  &  21.725   & 365 \\
   .. &   .....                 & .....   &   .....    &  .....  &   ..... &   .....                &   .....  &   .....   &   .....  & .....     & ....\\
  \\
\hline
\multicolumn{12}{l}{{ Note.} The entire table is available  in online
version. Only a portion of this table is shown here, to display its form
and contents.}\\
\end{tabular}
\label{table:sources}
}
\end{table*}

\section{Data and properties}
\label{s:data}

\subsection{Data sample}
\label{s:sample}
We have used SDSS DR12 quasar spectroscopic database compiled by
\citet[][]{Paris2017A&A...597A..79P}. This contains a
total of 297,301 spectroscopically confirmed quasars having spectra
covering the observed wavelength range of 3650-10400 \AA. Using this
catalog we have constructed our sample of quasar pairs by imposing
the following selection criteria: \ben
\item Consider only quasars with emission redshift $z>$ 2.57 to
  ensure that the wavelength range of \lya absorption is well covered
  by the SDSS spectrum. This condition is satisfied by 83,661 quasars.
  \item From the above 83,661 quasars, we have selected quasar pairs
    having angular separation $\theta <1.5$~\ar. This allows us to
    probe properties of the quasar environment over length scales of few
    100 \kpc. We identify 1,344 pairs satisfying this condition.
    \item Difference between the emission redshift of the background
      (i.e., $z_{b}$) and the foreground (i.e., $z_{f}$) quasar should
      be less than 0.5. This is to ensure that the \lya emission from
      the foreground quasar occurs in between the wavelength range of
      the \lyb and \lya emission of the background quasar. Only 380
      pairs out of the 1,344 aforementioned pairs satisfy this
      condition. \een \par In addition, we have also removed those
      quasars which show broad absorption line (BAL) features. The
      presence of broad absorption close to the broad emission lines
      could lead to large uncertainty in the estimation of the
      unabsorbed continuum flux and the emission redshift. From this
      list of 380 pairs, we have removed 68 pairs due to the presence
      of BAL feature in the background quasar and 42 pairs having BAL
      feature in their foreground quasar, based on the BAL flag in
      SDSS DR12 catalog. Furthermore, we also carried out a visual
      inspection of the remaining 270 pairs. This has lead us to the
      removal of an additional 5 pairs (one background and in four
      foreground quasars) that were not identified as BALs in the SDSS
      catalog. \par Similarly, the sightlines with very high H~{\sc i}
      column density absorption systems such as damped \lya absorption
      system (DLAs), sub-DLAs and Lyman-limit systems (LLS) in the
      proximity region are excluded from the analysis. For this
      purpose, we have used the catalog of DLAs/sub-DLAs by
      \citet[][]{Noterdaeme2009A&A...505.1087N,Noterdaeme2012A&A...547L...1N}
      to exclude the sightlines in which DLAs/sub-DLAs are found
      within 15 \Mpc~of the foreground quasar. This has resulted in
      the removal of 10 quasar pairs due to the presence of
      DLA/sub-DLA in background sightlines (within 15 \Mpc~of the
      foreground quasar). Two more quasar pairs were removed because
      of LLS ​falling in the ​transverse​ proximity region​ based on our
      visual inspection. Associated absorbers, ​if any, present within
      15 \Mpc~radial distance from the foreground quasar can give rise
      to an ​enhanced absorption, perhaps due to a ​possible​ outflow or
      inflow associated with them​. We found such associated absorption
      features in 1 background and 15 foreground sightlines. These 16
      pairs were removed from the sample and hence we are left with
      237 pairs in our sample.  ​Furthermore, one background quasar ​​is
      also found to show almost negligible continuum flux ​on either
      side of its strong nebular emission lines. This pair was also
      excluded as the \lya absorption cannot be probed in the
      transverse direction.
      \subsection{Emission redshift of the quasar}
      \label{s:zq}
      Estimation of accurate redshift for the quasars is important for
      the analysis of the proximity effect. In this context,
      \citet[][hereafter, HW10]{Hewett2010MNRAS.405.2302H} pointed out
      that the publicly available redshifts of SDSS DR7 quasars do
      possess systematic biases of $\Delta z/(1 + z) \ge$~0.002 i.e.,
      $\ge$~600 \kms~(blueward) which they have improved by a factor
      of $\sim 20$. The redshift measurement for quasars in our sample
      is based on improved SDSS DR12 pipeline using a linear
      combination of 4 eigen spectra
      \citep[e.g.,][]{SDSSpipeline2012AJ....144..144B}. To check the
      relative agreement between the redshift estimation based on SDSS
      DR12 pipeline with that obtained using HW10 algorithm (using
      C~{\sc iii}] emission line), we have used 21 quasars from our
        sample for which HW10 redshift measurements were also
        available. We found that the redshift measurements based on
        the algorithm used in SDSS DR12 and HW10 are consistent within
        1 $\sigma$ uncertainty of $\sim 30$ \kms~which is similar to
        the typical statistical redshift error quoted for quasars in
        the catalog of SDSS DR12. Nonetheless, we implement the HW10
        algorithm on our whole sample. The resulting redshift
        correction, based on C~{\sc iii}] emission line, is added to
          the original SDSS DR12 redshift which has lead to an
          improved redshift with negligible increase in their original
          statistical error, estimated by SDSS DR12 pipeline. \par
          Recently, \citet[][]{Shen2016ApJ...831....7S} have used
          Ca~{\sc ii} as the most reliable line for systemic redshift
          measurements and pointed out that the typical intrinsic and
          systematic uncertainty in the redshift measurement based on
          C~{\sc iii}] line is about 233 \kms~and 229
            \kms~respectively. As any such systematic redshift
              error in the emission redshift will be crucial for our
              analysis, therefore we have added the above systematic
              shift of 229 \kms~to our measured emission redshift. We
            have also added the above intrinsic and systematic
            uncertainties in quadrature with the statistical error
            ($\sim 30$ \kms~provided by SDSS pipeline) for each
            quasar. This results in a total redshift error estimate of
            $\sim 330$~\kms~which along with our corrected
              emission redshifts estimation are listed in
            Table~\ref{table:sources} for each quasar in our sample.

              \par In addition, we
            also visually checked the predicted emission line centroid
            for our entire quasar pairs sample, based on the redshift
            obtained using the above-mentioned procedures, for any
            visual abnormality (due to some possible poor
            characterization of the centroids). Here, we have been
            more stringent for any abnormality in the redshift
            estimation of the foreground quasars as compared to the
            background quasars, since our analysis will be more
            sensitive to any uncertainty in the former redshift. This
            has resulted in the removal of 29 quasar pairs (3
            background and 26 foreground sightlines). This criterion
            reduces the size of our sample to 207 quasar pairs. \par
            Finally, we demand that the velocity difference between
            the redshifts of the foreground and background quasars in
            a pair to be greater than 2000 \kms. This is required to
            distinguish the physically unassociated projected quasar
            pairs from the physically associated pairs. This leads to
            the removal of an additional 26 pairs. This resulted in
            our final sample of 181 pairs from SDSS DR12 for the
            analysis of the transverse proximity effect of the
            foreground quasars. \par From the above 181 quasar pairs,
            the spectra of the foreground quasars of each pair are
            used for the longitudinal proximity effect study. This has
            an additional advantage that we are probing the
            environment of the same set of quasars \citep[i.e., 181
              foreground quasars, see
              also][]{Kirkman2008MNRAS.391.1457K,Lu2011ApJ...736...49L}
            for our analysis of the transverse proximity effect
            (henceforth, TPE) as well as for the longitudinal
            proximity effect (henceforth, LPE).
 \subsection{Distances and Ly-continuum luminosity}
          \label{s:distance}
          For the analysis of TPE the proper radial distance ($r$)
          between the foreground quasar at $z_{f}$ and the absorbing
          cloud along the background quasar sightline pixel at
          $z^{b}_{a}$, is computed as
         \begin{equation}
           r= \sqrt{r^2_\perp(z_{f},\theta)+r^2_\parallel(z_{f},z^{b}_{a})}.
           \label{eq:prop_r}
         \end{equation}
         Here $r_\parallel$ is computed as
         \begin{equation}
           r_\parallel = \frac{c~\Delta~z}{(1+z_{f})~H(z_{f})}=\frac{\Delta~v}{H(z_{f})}
           \label{eq:r_para}
         \end{equation}
         where, $\Delta z = |z^{b}_{a}-z_{f}|$ is the redshift
         difference between the absorber along the background quasar
         sightline and the foreground quasar. Here, $H(z_{f}) =
         H_0\sqrt{\Omega_m(1+z_{f})^3+\Omega_\lambda}$ is the Hubble
         constant at $z_{f}$
         \citep[][]{Kirkman2008MNRAS.391.1457K}. We also note
           that in our above $r_\parallel$ estimation, we have ignored
           the effect of peculiar velocities, though it could be
           significant close to the quasar. Here, $r_\perp$ is the
         shortest distance in the plane of the sky from the background
         sightline to the foreground quasar (at $z_{f}$), which is
         computed as,
         \begin{equation}
           r_\perp = D_A(z_{f})\times\theta
           \label{eq:r_perp}
         \end{equation}
         where, $D_A(z_{f})$ is the angular diameter distance of the
         foreground quasar from the earth and $\theta$ (in radians) is the
         observed angular separation between the background and the
         foreground quasar sightlines
         \citep[][]{Hogg1999astro.ph..5116H}. Our sample has the
         perpendicular distances $(r_\perp)$ in the range of 25 to 700
         \kpc~with a median value of $\sim$ 566 \kpc~over the redshift
         range of 2.5 $\leq z \leq$ 3.5. For the distances used in the
         analysis of LPE, the proper distance between the foreground
         quasar at $z_{f}$ and the absorbing cloud pixel at
         $z^{f}_{a}$, is just the $r_\parallel$ by using
         Eq.~\ref{eq:r_para} (using $z^f_{a}$ instead of $z^b_{a}$).
         \par For the TPE analysis, we considered all the \lya forest
         pixels along the sightline of the background quasars
         corresponding to a proper radial distance (using
         Eq.~\ref{eq:prop_r}) smaller than 15 \Mpc~from the foreground
         quasar (the distance beyond which proximity effect can be
         safely ignored, e.g., Sect.~\ref{s:Analysis1}). At the same
         time, we also demand that these regions are at radial
         distances $r>10$ \Mpc~from the background quasars. This
         ensures that the region under consideration is not influenced
         by the ionizing radiation from the background quasar. For the
         analysis of LPE, the proximity region considered is just the
         \lya forest within $r_\parallel(z_{f},z^{f}_{a})< 15$
         \Mpc~along the line of sight to the foreground quasar and not
         influenced by the background quasar radiation.  \par Finally,
         the Lyman continuum luminosity of the quasar at 912
         \AA~($L_{\nu_{912}}$) is estimated as,
            \begin{equation}
              L_{\nu_{912}} = F_{\nu_{912}} \times~4\pi~d_L^2
               \label{eq:lnu_fnu}
            \end{equation}
              where $d_L$ is the luminosity distance from the observer to
              the quasar. The flux $ F_{\nu_{912}}$ is estimated from
              the observed flux at the closest line free region
              to the \lya forest around 1325 \AA~in the
              quasar's rest frame and then extrapolated to 912 \AA~by
              using broken power law of the form,
       \begin{equation}
         F_\nu \propto \biggl\{\begin{array}{ll}\nu^{-\alpha}~~;~\lambda > 1300~\AA
         \\   \nu^{-\alpha_{uv}}~~;~\lambda < 1300~\AA \end{array}
         \label{eq:power}
       \end{equation}
              with $\alpha=$ 0.44 and $\alpha_{uv}=$ 1.57 \citep[e.g., see][and
                references therein] {khaire2015ApJ...805...33K}.
                Here, we have assumed a very simplistic treatment of
                the quasar spectral energy distribution (i.e., a constant
                value of $\alpha_{uv}$ for all quasars). However, the
                effect of scatter in $\alpha_{uv}$ is found to be
                negligible in our final analysis (as discussed in
                Sect.~\ref{s:alpha_var}). Some of the properties of
              the 181 quasar pairs in our sample are plotted in
              Fig.~\ref{Fig:qso_prop}. In the top panel, we show a
              plot of the $L_{\nu_{912}}$ of the background quasars
              versus $z_{b}$. The same plot for the foreground quasars
              is shown in the middle panel. In the bottom panel, we
              plot the $L_{\nu_{912}}$ of the foreground quasars
              versus $r_{\perp}$. The top and right histograms
              in each plot show the distribution of the parameters labelled on
               the abscissa and ordinate respectively. The figure also
              displays quasar pairs in our sample which are found to
              be common with that of
              \citet[][]{Prochaska2013ApJ...776..136P}. The specific
              luminosity at 912 \AA~of their whole sample ranges from
              $10^{29.5}$ to $10^{31.2}$ erg s$^{-1}$ Hz$^{-1}$ and
              $r_\perp$ ranges from 30 kpc to 1 \Mpc. The details of
              the 181 quasar pairs used in our analysis are listed in
              Table~\ref{table:sources}.
              \begin{figure*}	
                \centering
                \includegraphics[height=8cm,width=8.0cm,trim=0cm 0 0 0]{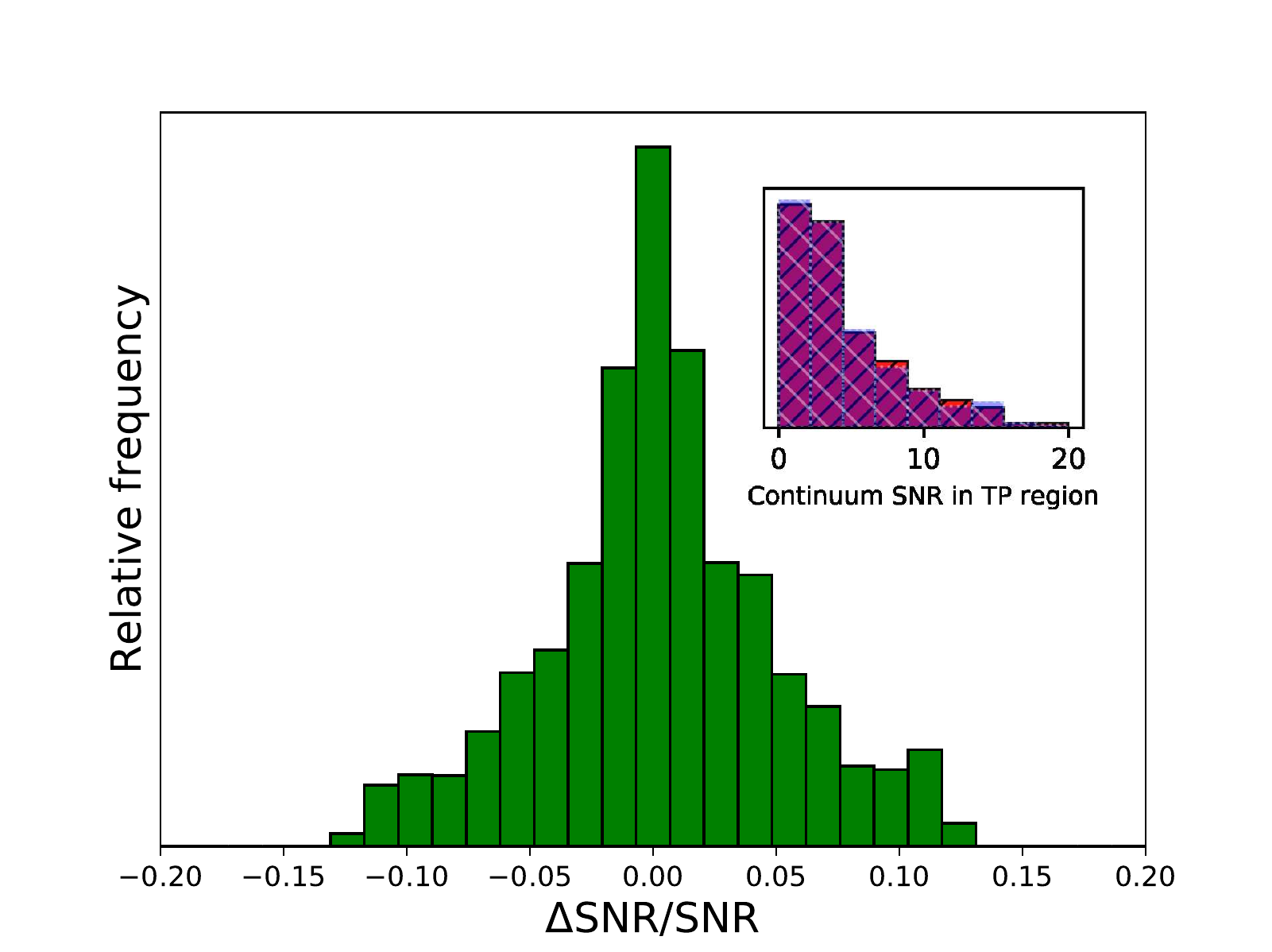}
                \includegraphics[height=8cm,width=8.0cm,trim=0cm 0 0 0]{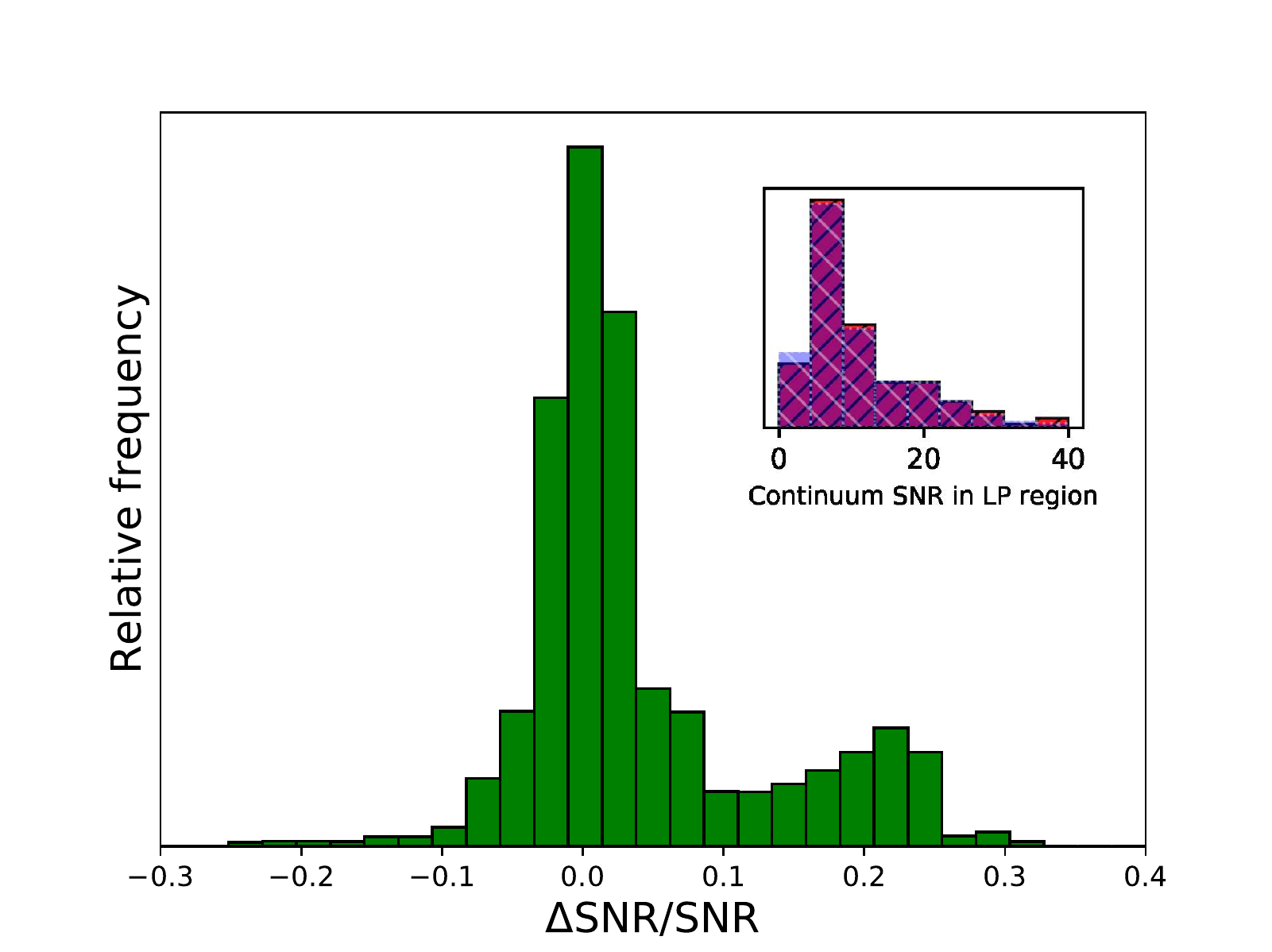}
                \caption {{\it Left:} The histogram plot showing the
                  distribution of $\Delta$SNR/SNR ($\equiv$
                  [SNR$_{prox}$-SNR$_{\textsc{igm}}$]/SNR$_{prox}$)
                  based on the SNR of the spectra of the
                  background quasars (SNR$_{prox}$) and the redshift
                  matched control sample (SNR$_{\textsc{igm}}$). {\it
                    Right:} Same as left but from the spectra of the
                  foreground quasars and their corresponding control
                  sample. In the insets, we plot the SNRs of the
                  spectra of quasars in our main sample
                  (white-slanted) and its corresponding control sample
                  (black-slanted) in the proximity region used in our
                  analysis.}
               \label{Fig:z_snr_match}
             \end{figure*}
 \subsection{Control sample of \lya forest}
 \label{s:control_sample}
 For carrying out the statistical analysis of the \lya absorption, we
 use pixel optical depth, $\tau(\lambda_i)$, statistics
 \citep[e.g.,][]{Rollinde2005MNRAS.361.1015R,Aglio2010ApJ...722..699D},
 where, $\tau(\lambda_i)=-ln[F(\lambda_i)/F_{c}(\lambda_i)]$ is the
 effective optical depth integrated over the pixel width (69 \kms~for
 SDSS spectra) of the observed spectrum. Here, $F(\lambda_i)$ and
 $F_{c}(\lambda_i)$ are the observed flux and the unabsorbed continuum
 flux respectively at the $i^{th}$ pixel having a wavelength
 $\lambda_i$. In this method, the proximity effect analysis is carried
 out by performing a statistical comparison of a probability
 distribution of pixel optical depth obtained from the proximity
 region with that originating from the general IGM. However, while
 combining the optical depths of \lya absorption from different
 redshifts it is important to account for the strong redshift
 evolution of the optical depth.  \par One possibility is to scale the
 optical depth measured at various redshifts to their expected value
 at some fixed reference redshift \citep[e.g.,
   see][]{Rollinde2005MNRAS.361.1015R,Agliolpe32008AA...480..359D,Kirkman2008MNRAS.391.1457K,Faucher2008ApJ...681..831F}.
 However, given the fact that the observed optical depth distribution
 has lower and upper limits based on the noise in the continuum and
 the core of the saturated absorption lines, such scaling may
 artificially introduce the pixel optical depth values beyond these
 observational limits. This may bias the statistics by creating an
 artificial difference between the pixel optical depth distributions
 measured for the IGM and the proximity region. This issue is
 particularly important for our sample as it is based on the SDSS
 spectra which are obtained with a low/moderate resolution ($R\sim
 2000$) and low SNR (mostly $<20$) where the observed line profiles
 can easily hide the saturated absorption lines.

  \par An alternate approach to the redshift scaling is to use a
  control sample of \lya absorption from the IGM. A control sample
  spectra matching closely in the redshift and in the continuum SNR
  with that of the spectra of the quasars in our pair sample can be
  used for studying the proximity effect. Therefore, even though
    there is redshift evolution of optical depth measured in the
    proximity regions, a similar evolution will also be present in the
    control sample of the IGM, which will be used as a comparison
    sample. We adopt this approach in our analysis. For
    constructing such a control sample, we made use of the non-BAL
    quasar catalog from SDSS DR12 after cross-correlating with the
    DLAs/sub-DLAs catalog of
    \citet[][]{Noterdaeme2009A&A...505.1087N,Noterdaeme2012A&A...547L...1N}
    to remove any sightlines with DLA. This sample of quasars without
    BAL and DLA along their line-of-sight will be referred as IGM
    parent sample. \par To choose SNR matched control sample from this
    IGM parent sample, consisting of $\sim$ 63,000 quasars, we
    calculated a running median SNR
    (continuum/$\langle$error$\rangle$) over a window of 75 pixels in
    the \lya forest. These 75 pixels correspond to a radial distance
    of $\sim$ 15 Mpc at a redshift of $z\sim$ 3 for a spectrum
    observed with SDSS resolution. The control sample for each
    sightline of our 181 quasar pairs were generated from the above
    IGM parent sample, such that $\Delta$SNR/SNR ($\equiv$
  [SNR$_{prox}$-SNR$_{\textsc{igm}}$]/SNR$_{prox}$\footnote{
    Throughout the paper we will use ``prox'' and ``IGM'' as
    sub-scripts to represent the 181 quasar pair sample and control
    sample respectively.}) is as small as possible in the spectral
  range corresponding to the 15 Mpc of the proximity region. Moreover,
  for reasonably good statistics, we demand 25 quasars as a control
  sample for each quasar in our pair sample. With these criteria, it
  can be noted from Fig.~\ref{Fig:z_snr_match} that we could achieve
  $\Delta$SNR/SNR less than $\pm$0.15 for the TPE analysis (using
  background quasar's sightlines) and $\pm$0.35 for the LPE analysis
  (using foreground quasar's sightlines). In the LPE analysis, due to
  the presence of strong \lya emission line, the SNR in the spectral
  region used for longitudinal proximity analysis is typically
  high. Therefore, due to the scarcity of high SNR SDSS spectra, we
  have to allow a larger window of $\Delta$SNR/SNR (i.e., $\pm$
  0.35). We have also ensured non-repetition of sightlines while
  constructing the control sample for each direction, to remove the
  possible significant contribution from repeated use of any peculiar
  sightline. \par
\begin{figure*}	
   \centering
   \includegraphics[height=12cm,width=14.0cm,scale=0.5,trim={0cm 0cm 0cm 0cm}]{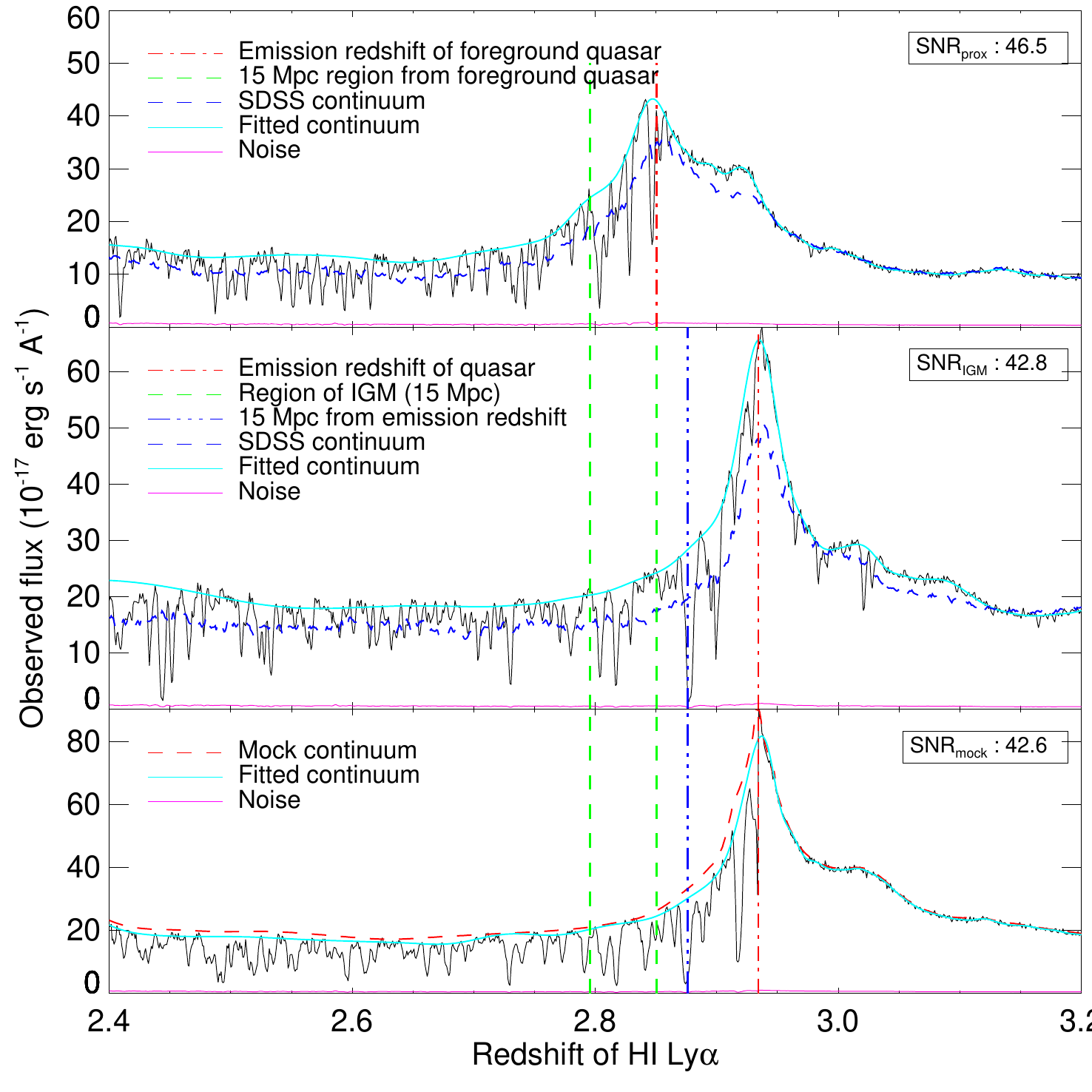}
   \caption {{\it Upper panel: } SDSS spectrum of one of the
     foreground quasars (J2156+0037) at $z=$ 2.85 (red, dot-dashed
     vertical line), which has the highest SNR in our sample of
       the foreground quasars overlaid with the SDSS continuum (blue,
     dashed) along with our improved continuum, fitted using iterative
     B-spline fitting method (cyan, solid). The vertical green dashed
     lines represents the proximity region (15 Mpc) considered in our
     analysis. The noise spectra is represented with magenta
       color along with the median SNR (in the proximity region)
       stated at the upper right corner.  {\it Middle panel: } Same
     as the upper panel but for one of the quasars in the
     corresponding control sample of the above foreground quasar. The
     red dot-dashed line shows the emission redshift of the control
     sample while the blue dot-dot-dot-dashed line shows its proximity
     region. {\it Lower panel: } The plot shows the mock spectrum
       of the quasar used as control sample above, overlaid with the
       true continuum (red, dashed) along with our fitted continuum
       (cyan, solid).}
   \label{Fig:conti_max}
    \end{figure*}

\begin{figure*}	
     \centering
     \includegraphics[height=12cm,width=14.0cm,scale=0.5,trim={0cm 0cm 0cm 0cm}]{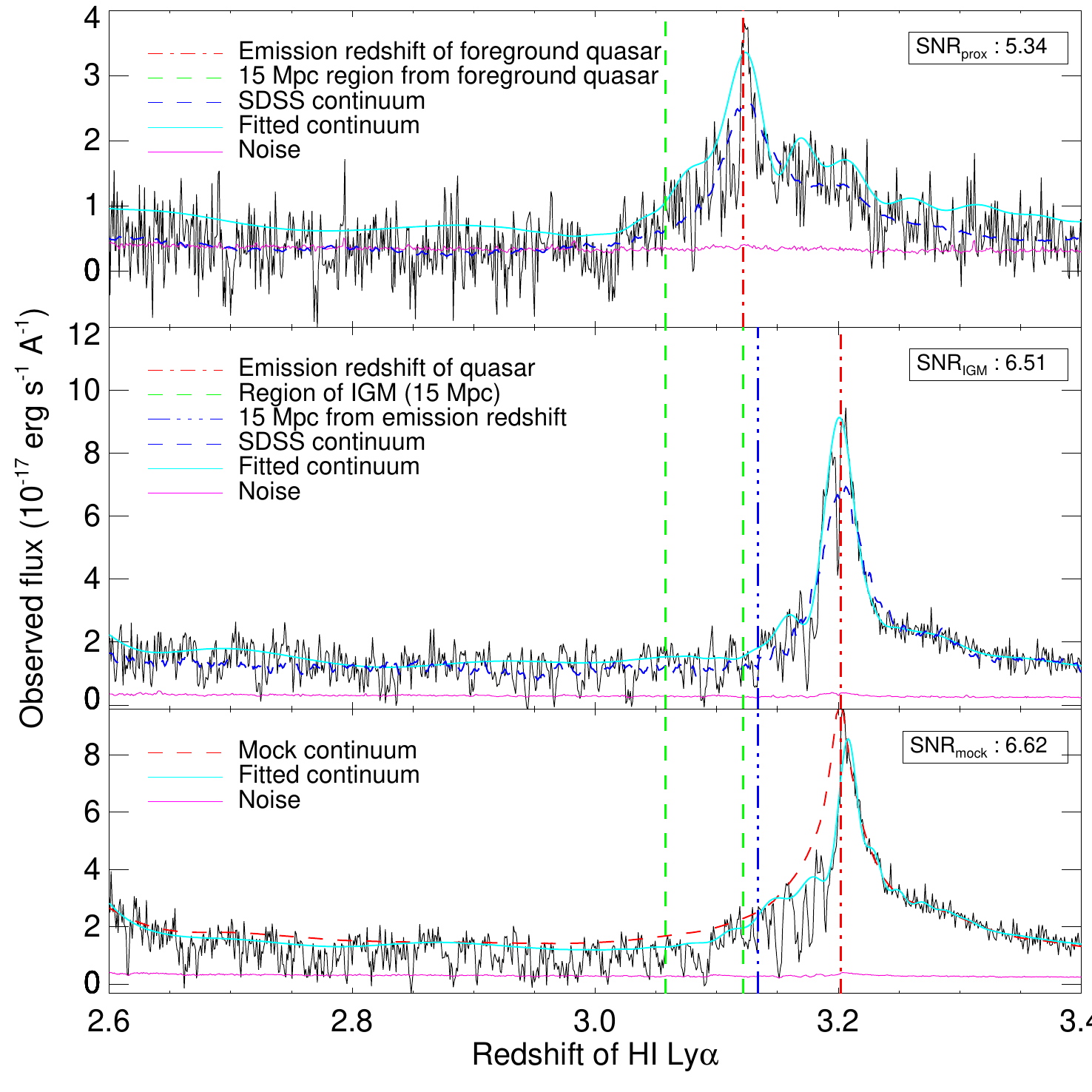}
     \caption {Same as Fig.~\ref{Fig:conti_max} but illustrated here
       by taking an example of a low SNR foreground quasar in our
       sample (J1111+0613 at $z=$ 3.12).}
   \label{Fig:conti_min}
 \end{figure*}

 Furthermore, we checked all the quasar spectra in
  the control sample, visually, for any peculiarity and LLS in the
  wavelength region corresponding to the proximity region of the
  quasar in the 181 quasar pairs sample. This led to the exclusion of
  just 10 and 7 sightlines from the control sample constructed for TPE
  and LPE respectively. Using this procedure, we have ensured the
  exact redshift matching along with a good matching of the continuum
  SNR between the spectral range used for the proximity analysis and
  the corresponding spectral range in the control sample (e.g., see
  Fig.~\ref{Fig:z_snr_match}). We would like to point out the fact
  that the IGM region from the control sample is required to be at
  least 15 Mpc from its emission redshift to avoid its longitudinal
  proximity region. Therefore, the emission redshift of the control
  sample may vary from its corresponding proximity sample. In the
  inset of the Fig.~\ref{Fig:z_snr_match}, we have shown the SNR of
  the proximity sightlines and control sample sightlines over the
  wavelength range used in our analysis. In the upper two panels of
  Fig.~\ref{Fig:conti_max}, we have shown an example of our high SNR
  proximity sample and one of its corresponding control sample. A
  similar plot for the low SNR sightline is shown in
  Fig.~\ref{Fig:conti_min}. Additionally, we note that the quasars in
  both the comparison samples in our analysis, viz. pairs sample and
  the corresponding control sample, were observed in the SDSS using
  the same spectral setting, as a result, any effect of spectral
  resolution will also be similar for both of them.

   \section{Analysis and results}
   \label{s:Analysis}

  \subsection{Flux and optical depth analysis}
  The normalized transmitted flux $[F_{t}(\lambda_i)]$ of a quasar
  spectrum at i$^{th}$ pixel having wavelength $\lambda_i$ is given
  by,
  \begin{equation}
    F_{t}(\lambda_i) \equiv F(\lambda_i)/F_{c}(\lambda_i) =
    e^{-\tau(\lambda_i)}
    \label{eq:ft}
  \end{equation}
     where, $F_{c}(\lambda_i)$ is the unabsorbed continuum flux fitted
     to the observed flux $F(\lambda_i)$ and $\tau(\lambda_i)$ is the
     pixel optical depth. We compare the distribution of the
     normalized transmitted flux as well as of the pixel optical depth
     measured in the proximity region to that in the SNR and redshift
     matched spectral region along the line-of-sight to the quasars in
     the control sample. The continuum placement in the \lya forest
     and its associated uncertainty can have a significant impact on
     both the transmitted flux and the optical depth statistics, which
     we have investigated with a detailed analysis in the next
     subsection.

 \subsubsection{Quasar continuum normalization and it's uncertainty}
\label{s:conti}
During our visual inspection, we noticed that the continuum fit given
by SDSS DR12 pipeline systematically underestimates the continuum flux
especially in the \lya forest as evident from top panels of
Fig.~\ref{Fig:conti_max} and Fig.~\ref{Fig:conti_min} for it's
illustration in a high and low SNR spectrum respectively \citep[see
  also][]{Lee2012AJ....143...51L}. Therefore, we have refitted the
continuum for all quasar spectra used in our study. For this, we have
used iterative B-spline fitting along with the median smoothing
function by optimizing the fitting parameters interactively to get the
best continuum fit. The procedure adopted here is similar to that
  of \citet[][]{Aglio2008AA...491..465D,Agliolpe32008AA...480..359D}.
As a first step, we divide the whole spectrum into small intervals
such that the emission line region has a smaller interval length to
take into account a large flux gradient over this region. Each of
these small chunks of wavelength versus flux are then fitted by an
iterative B-spline fitting algorithm. The order of the fitted B-spline
does vary among various quasar spectra but in most of the cases, a
$4^{th}$ order has resulted in an optimal fit. Secondly, we minimize
the $\chi^2$ between the flux spectrum and our modelled continuum in
each chunk of the spectrum iteratively. In each iteration, a certain
number of pixels got rejected which lies outside the range of our
allowed 3$\sigma$ significance level, leading to the removal of high
absorption pixels iteratively in each chunk. The full continuum is
then the smoothed function of the fitted continuum to these chunks.
This leads to an improved continuum fit to all quasar spectra used in
this study as compared to the default SDSS continuum as shown in the
top panels of Fig.~\ref{Fig:conti_max} (e.g., for high SNR spectrum)
and Fig.~\ref{Fig:conti_min} (e.g., for low SNR spectrum).

  \begin{figure*}   
  \centering
  \includegraphics[height=8cm,width=8cm,scale=0.6,trim={1cm 0cm 0cm 0cm}]{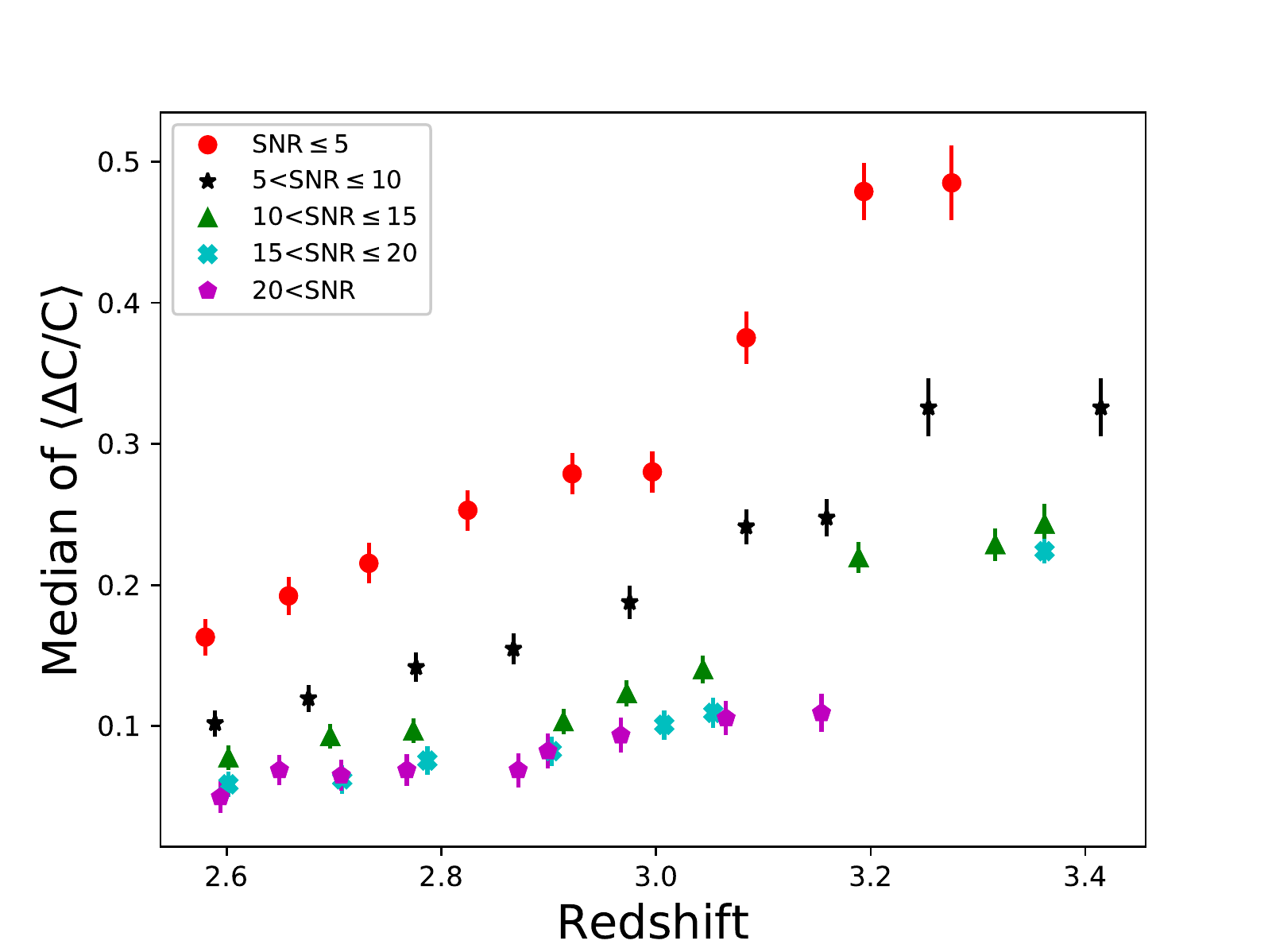}
  \includegraphics[height=8cm,width=8cm,scale=0.6,trim={1cm 0cm 0cm 0cm}]{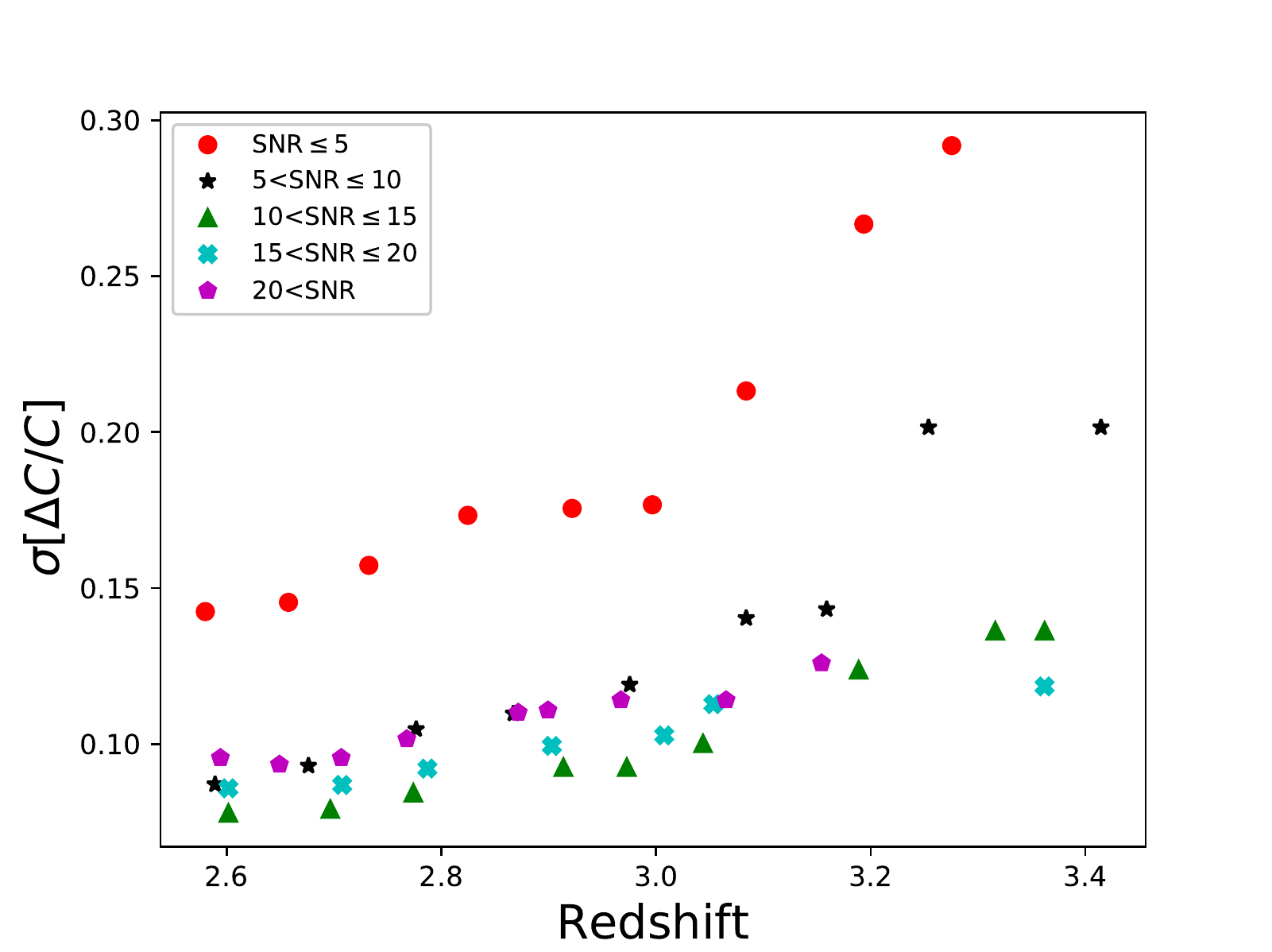}
  \caption {{\it Left panel:} The plot shows median value of continuum
    shifts $\langle \Delta C/C \rangle$ versus redshift measured in
    various SNR bins using the mock spectra from
    \citet[][]{Bautista2015JCAP...05..060B}. The error bars on the
    mean shift plotted here are enhanced by a factor of 9, which
    otherwise are smaller than the symbol size.  We have incorporated
    such continuum shift in our analysis (e.g., see
    Sect.~\ref{s:conti}). It can be seen from the figure that as
    expected based on the redshift evolution of \lya optical depth,
    the $\langle \Delta C/C \rangle$ at a given SNR increases with
    increasing $z$. The effect seems stronger for the lower SNR.  {\it
      Right panel:} The plot shows the median value of the fractional
    uncertainty in the continuum ($\sigma [\Delta \textsc{c/c}]$)
    versus the redshift measured in the various SNR bins using the
    mock spectra from \citet[][]{Bautista2015JCAP...05..060B}.}
      \label{Fig:delc_z_snr}
  \end{figure*}

\par However, due to numerous absorption lines in the \lya forest, a
systematic continuum placement error may still remain (particularly
for low SNR spectra). To quantify this, we have used the mock quasar
spectra simulated by \citet[][]{Bautista2015JCAP...05..060B}. They
have obtained the mock spectrum in three steps: Firstly, they
generated the \lya optical depth values along the line-of-sight. Then
this simulated normalized flux ($F=e^{-\tau}$) is multiplied by a
synthetic quasar continuum flux ($F_c$). Secondly, the mock spectra are
convolved with the kernel to have a similar resolution ($R\sim
2000$). Furthermore, they have also ensured that mock quasar has
emission redshift, the mean flux in the \lya forest and a spectral index
similar to those of the corresponding real quasar found in SDSS DR11
quasar catalog. Finally, instrumental noise, metal lines, high column
density absorbers and other potential sources of systematics are
added to the spectrum to match statistically with the real SDSS quasar
spectrum. The noise added to the fluxes of a given mock quasar is a
random number from a Gaussian distribution of mean zero with a
variance determined by the noise model for the corresponding real
quasar. In the bottom panels of Fig.~\ref{Fig:conti_max} and
Fig.~\ref{Fig:conti_min}, we have shown the mock spectrum for the
corresponding IGM control sample plotted in the middle panels. For
each real spectrum in our sample, we took 100 such realizations of its
mock spectra from \citet[][]{Bautista2015JCAP...05..060B}.\par We
fitted the continuum to all these mock spectra using our continuum
fitting procedure.  This allows us to compute the fractional
uncertainty between the true ($C_t$, red-dashed line of bottom panels
of Figs.~\ref{Fig:conti_max}, \ref{Fig:conti_min}) and the fitted
continuum ($C_f$, cyan-solid line of bottom panels of
Figs.~\ref{Fig:conti_max}, \ref{Fig:conti_min}) over the wavelength
range relevant for our proximity analysis (i.e., within vertical
green-dashed line of Figs.~\ref{Fig:conti_max}, \ref{Fig:conti_min}),
viz., $\Delta C/C~\equiv~[C_t-C_f]/C_f$, for each pixel. The
mean of this distribution ($\langle\Delta C/C\rangle$) is used to
apply a systematic continuum shift in all the pixels of the proximity
region to obtain the unabsorbed intrinsic continuum ($F_c[\lambda_i]$)
as,
  \begin{equation}
    F_c[\lambda_i] =  C_f[\lambda_i]\times(1+\langle\Delta C/C\rangle).
    \label{eq:fc_lam}
  \end{equation}
  The typical uncertainty in the continuum flux is computed using the
  standard deviation ($\sigma [\Delta \textsc{c/c}]$) of this $\Delta
  C/C$ distribution as follows,
  \begin{equation}
   \delta F_c[\lambda_i] = \sigma[\Delta\textsc{c/c}]C_f[\lambda_i]\times \Big(1+\langle\Delta C/C\rangle\Big)
  \label{eq:err_fc}
  \end{equation}
 Here, we have neglected the
    error on the mean value of $\Delta C/C$ as it is found to be
    negligible compared to the above systematic continuum error.

The mock spectra were available for about 67\% of the total quasar
sightlines used in our analysis. The unavailability of the mock
  spectra for the remaining 33\% sightlines is probably due to the
  fact that \citet[][]{Bautista2015JCAP...05..060B} has confined their
  study only to the quasars catalog of SDSS DR11. However, our sample
  (both quasar pairs and control sample) is based on SDSS DR12 quasars
  catalog which has added a substantial number of new quasars beside
  including the quasars from SDSS DR11 quasars catalog. For these
remaining 33\% quasar sightlines without the mock spectra, we have
applied the continuum correction by using the shift based on those
mock spectra which are closely matching in the SNR and the
redshift. \par The fractional averaged systematic continuum shift (i.e.,
$\langle\Delta C/C\rangle$) and the typical continuum uncertainty per
pixel as a function of redshift is plotted in the left and right panel of
Fig.~\ref{Fig:delc_z_snr} respectively, for various SNR bins.  As can
be seen from the figure that, (i) as the SNR increases, the fractional
systematic continuum shift decreases and (ii) there seem to be a
moderate increasing trend with the redshift for higher SNR bins in
comparison to a strong increasing trend apparent for lower SNR
bins. The typical continuum uncertainty also follows similar
  trend except for very high SNR bin. In very high SNR bin, our
  continuum procedure optimized for moderate SNR spectra may start
  tracing the absorption, leading to higher uncertainty. This shows
the importance of applying the continuum correction along with
  the proper estimation of its uncertainty computed for individual
sightline. In our analysis that follows, we use the corrected
continuum flux (e.g., see, Eq.~\ref{eq:fc_lam}) for each sightline
along with the error spectrum (e.g., see, Eq.~\ref{eq:err_fc}). This
will allow us to include the continuum placement uncertainties in the
normalized transmitted flux (e.g., see, Eq.~\ref{eq:ft}) along with
the other measured flux uncertainties at each pixel as discussed
  in the next subsection.
    \begin{figure*}   
               \centering
               \includegraphics[height=8.7cm,width=8.9cm]{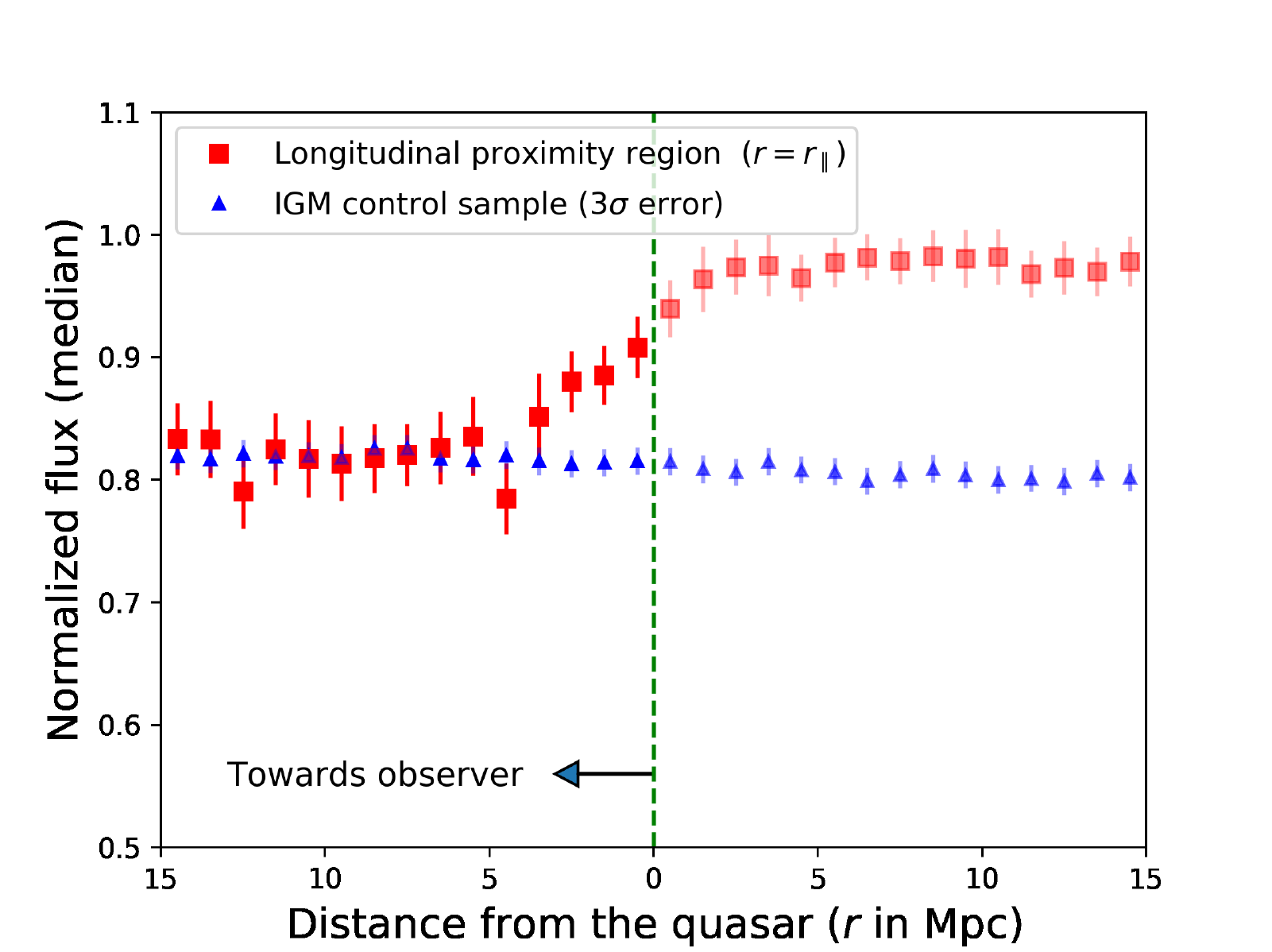}
               \includegraphics[height=8.7cm,width=8.9cm]{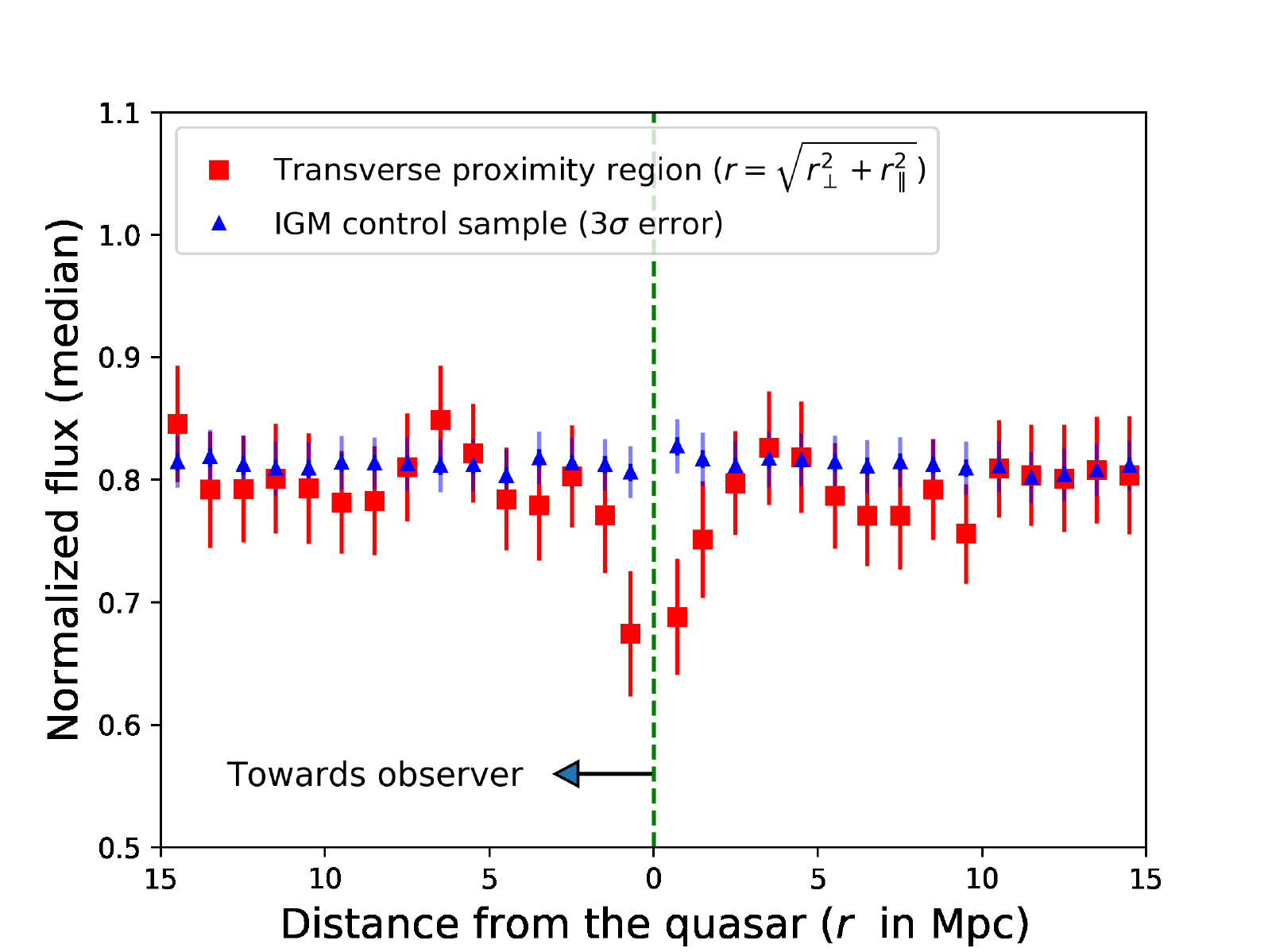}
               \caption{The plots show median transmitted flux in
                 different radial distance bins from the foreground
                 quasars for the longitudinal (left) and the
                 transverse proximity regions (right). The plots also
                 show the corresponding median transmitted flux for
                 the control sample (blue, triangle) selected based on
                 appropriate matching of the redshift and SNR in
                 the proximity region (i.e., LPE and TPE of our
                 foreground quasars) of our main sample (e.g., see
                 Sect.~\ref{s:control_sample}). For the LPE (left
                 panel) it may be noted that the quasar's continuum
                 flux beyond \lya emission line is
                 shown in fainter color to distinguish them from real
                 \lya absorption. In this plot, the error bars on
                 the median transmitted flux of the proximity region
                 consists of flux error from photon counting
                 statistics, error due to the continuum placement
                 uncertainties, r.m.s statistical errors,
                 sightline-to-sightline variance and emission redshift
                 measurement error within the 1 \Mpc~radial distance
                 bin (e.g., Sect.~\ref{s:Uncertainties}), whereas for
                 control sample only the former three error
                 contributes. Here, the error-bars in the control
                 sample are smaller than the symbol size due to
                 statistics of large numbers (being $\sim$ 25 IGM
                 sightlines per quasar sightline).  Therefore, we have
                 plotted the 3$\sigma$ error for the IGM median
                 flux. The green dashed line shows the location of the
                 foreground quasar.}
               \label{Fig:rad_flux_red}
  \end{figure*}
  \begin{figure*}   
               \centering
               \includegraphics[height=8.7cm,width=8.9cm]{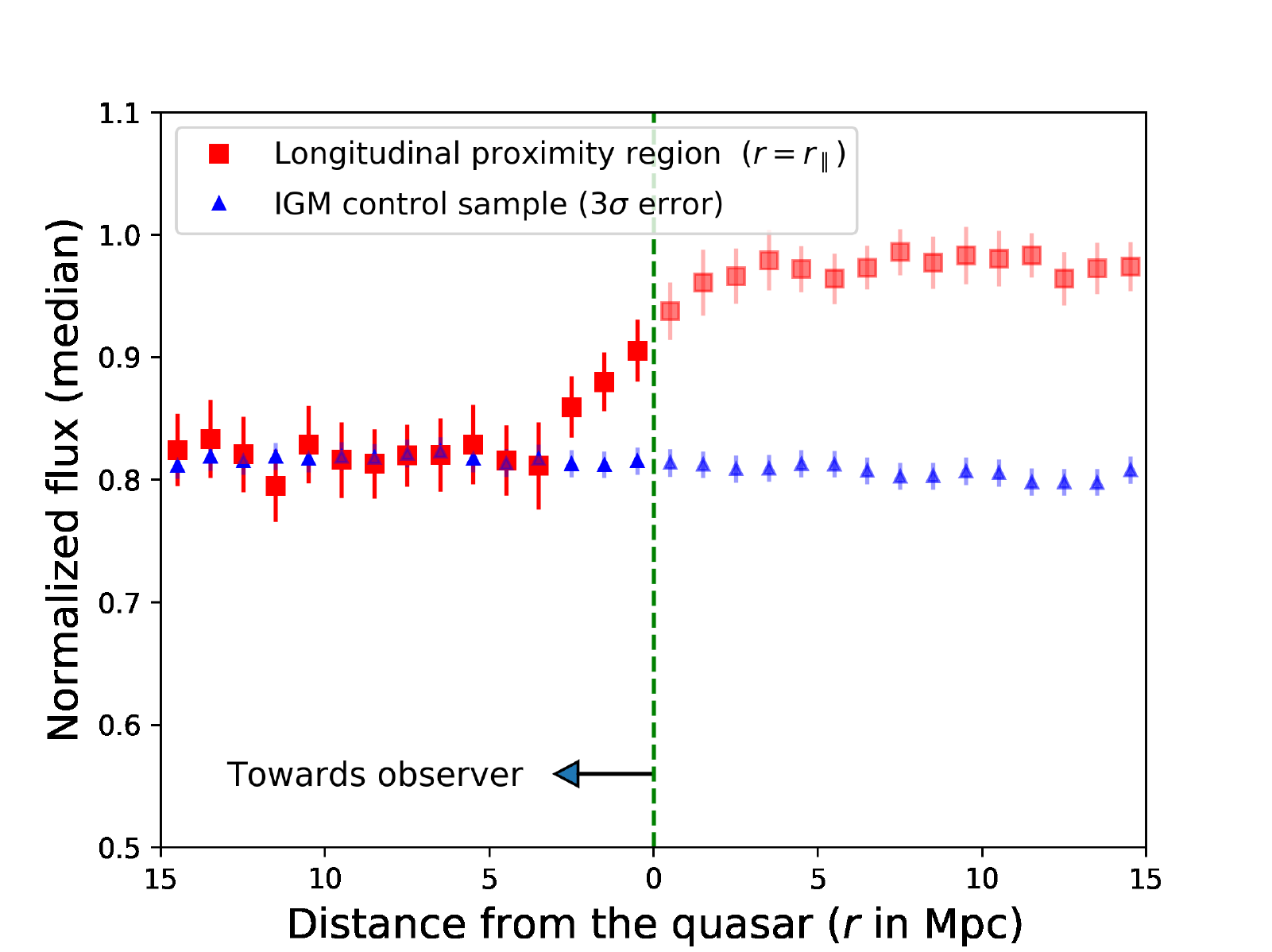}
               \includegraphics[height=8.7cm,width=8.9cm]{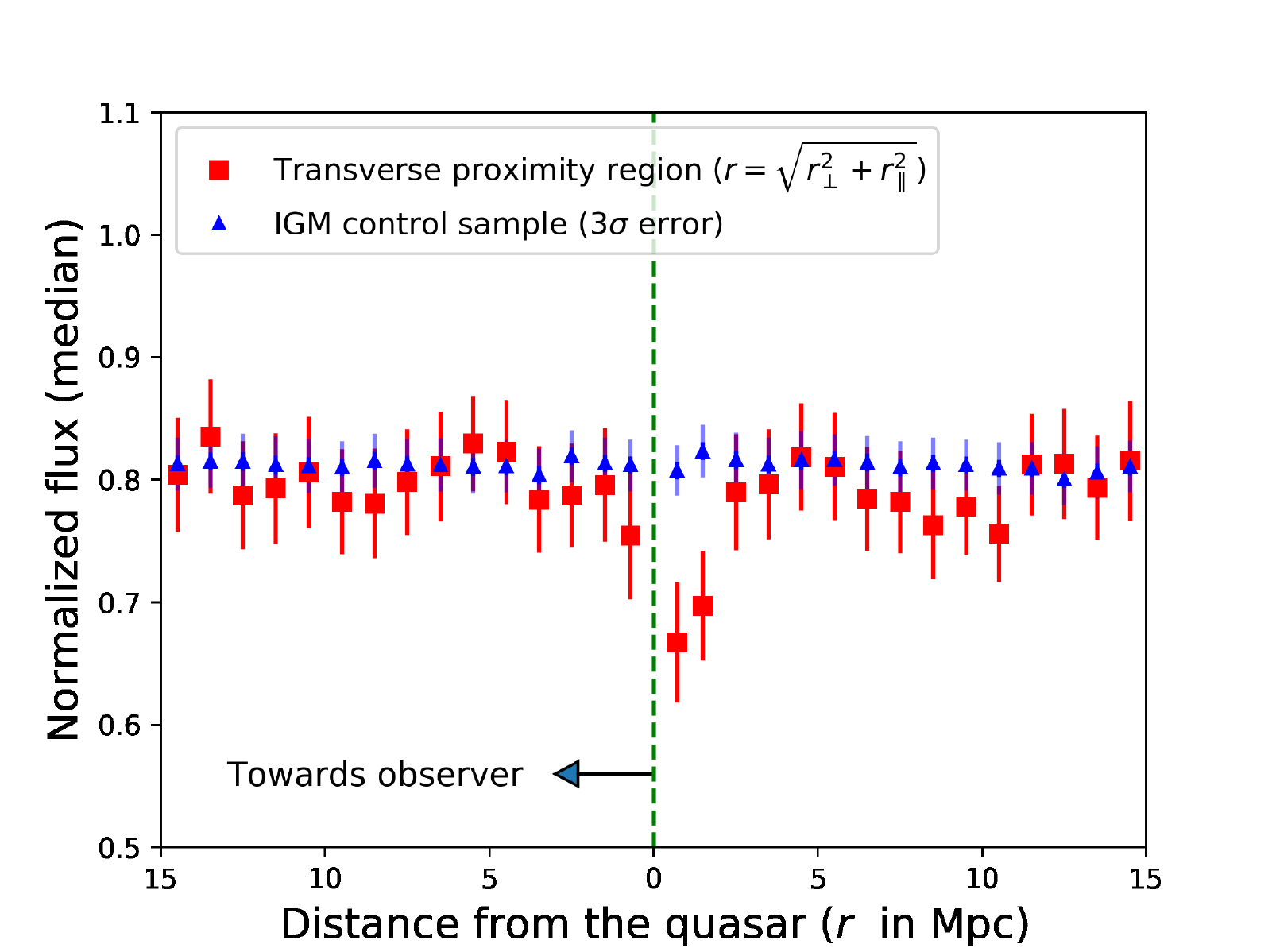}
                  \caption{Same as Fig.~\ref{Fig:rad_flux_red} except
                    that here we have not applied the systematic
                    redshift uncertainty correction of 229 \kms~to the
                    emission redshift of the foreground quasar (e.g.,
                    see Sect.~\ref{s:zq}) leading to an asymmetry
                    among blueward and redward of the foreground quasar as
                    evident in the transverse direction.}
               \label{Fig:rad_flux}
  \end{figure*}

 \subsubsection{ Transmitted flux uncertainties}
\label{s:Uncertainties}
      To quantify the difference in $F_t$ (e.g., see, Eq.~\ref{eq:ft})
      for a given radial distance bin of the proximity region as
      compared to its control sample, it is important to carefully
      consider all the probable sources of uncertainties involved in
      the measurement of the median $F_t$. In our analysis, the
        first contribution of error ($\Delta F^{fc}_t$) in the
        normalized spectrum comes from the propagation of flux
        measurement errors and continuum placement uncertainties at
        each pixel, which is computed as follows,
  \begin{equation*}
    \Delta F^{fc}_t[\lambda_i]  = F_t[\lambda_i] \times
    \sqrt{\Bigg(\frac{\delta
        F[\lambda_i]}{F[\lambda_i]}\Bigg)^2+\Bigg(\frac{\delta
        F_c[\lambda_i]}{F_c[\lambda_i]}\Bigg)^2}. 
 \end{equation*}
    Here, $\delta F_c[\lambda_i]$ is the error introduced due to the
    continuum fitting uncertainties derived using mock spectra (e.g.,
    see Eq.~\ref{eq:err_fc}) for a given $F_c[\lambda_i]$ (e.g., see
    Eq.~\ref{eq:fc_lam}) at the $i^{th}$ pixel with wavelength
    $\lambda_i$. $\delta F[\lambda_i]$ is the error on the flux
    measurement $F[\lambda_i]$ provided by the SDSS
    pipeline. Therefore, the error on the mean transmitted flux (from
    all the 181 spectra) in a radial distance bin is obtained using,

      \begin{multline}
\Delta F^{fc}_t 
    = \Bigg[\frac{\sum_{j=1}^{N_{spec}}\sum_{i=1}^{N_j}
       F^j_t[\lambda_i]^2 \times \Big(\frac{\delta
         F^j[\lambda_i]}{F^j[\lambda_i]}\Big)^2}{(\sum_{j=1}^{N_{spec}}N^j)^2} \\
     + \frac{\sum_{j=1}^{N_{spec}}\sum_{i=1}^{N_j}
       F^j_t[\lambda_i]^2 \times \Big(\frac{\delta
         F_c^j[\lambda_i]}{F_c^j[\lambda_i]}\Big)^2}{(N_{spec})^2}\Bigg]^{1/2}
    \label{eq:f_er_p}
 \end{multline}
 The first and the second term are the photon counting noise and
   the continuum placement error respectively. N$_{spec}$ is the
   number of spectra i.e., 181 and N$_j$ is the number of the pixels
   in the j$^{th}$ spectrum used within the radial distance bin of
   size 1 Mpc (typically 4-6 pixels). Here, it may be noted that the
   continuum placement error in a given spectrum might be correlated.
   We have taken this into account by averaging it over only the number of
   spectra (expected to be uncorrelated) instead of averaging it over
   the total number of the pixels in a given radial distance bin
   (e.g., see the second term). 

 \par Second
 contribution to the error in the normalized flux is due to the
 dispersion of $F_t$ among the various pixels around its median value
 in a given radial distance bin (i.e., r.m.s. scatter, $\Delta
 F^{rms}_t$). This r.m.s. scatter is calculated as
  \begin{equation}
 \Delta F^{rms}_t = \frac{\sqrt{\sum_{i=0}^{N}[{F_t(i)-\langle F_t
         \rangle}]^2}}{N}.
 \label{eq:f_er_rms}
  \end{equation}
 Here, $N=\sum_{j=1}^{N_{spec}}{N_j}$, is the total number of
    the pixels in 1 Mpc radial bin. \par Third contribution to the total
  error budget is due to the sightline-to-sightline variance of the
  sample ($\Delta F^{var}_t$). For this, we have used the empirical
  bootstrap technique \citep[e.g., see][]{Efron1993}. In this method,
  we constructed a new sample of 181 quasars proximity sightlines by
  randomly selecting them from our original dataset of 181 quasars
  (i.e., allowing a random exclusion of sightlines at the cost of the
  equal number of random repetition of some other sightlines). The
  histogram of the median transmitted flux of 100 such realizations,
  within the spectral range corresponding to 1 Mpc spatial separation
  (i.e., radial bins corresponding to the proximity region), are well
  fitted with a Gaussian profile. This results in an average standard
  deviation in the transmitted flux of $\sim$ 2\% per 1 \Mpc~radial
  distance bin (varying within the range of $1.0\%$ to $3.0\%$ in
  different radial distance bins). This standard deviation has been
  used to include the sightline-to-sightline variance in our final
  error budget. \par Lastly, we also included the uncertainty in the
  median $F_t$ measured in the proximity region due to the typical
  emission redshift uncertainty ($\Delta F^{z}_t$) of $\sim$ 330
  \kms~along our sightlines as discussed in Sect.~\ref{s:zq}. For
  this, we have carried out an analysis of 100 realizations by adding
  a random velocity offset (with Gaussian distribution) to the
  individual quasar redshifts within $\pm$330 \kms~range. This offset
  in the redshift will propagate by affecting the inferred distance
  between the quasar and the absorber in the form of the number of
  pixels to be considered in a given radial distance bin. As a result,
  the corresponding uncertainty in the $F_t$ is estimated based on the
  spread of the measured median $F_t$ among these 100 realizations,
  for each radial distance bin. \par Therefore, the total error on the
  transmitted flux of the proximity region ($\Delta F^{prox}_t$) for a
  given radial distance bin is the quadratic sum of all the
  above-mentioned errors (e.g., see
  Eqs.~\ref{eq:err_fc},~\ref{eq:f_er_p} and~\ref{eq:f_er_rms}),
 \begin{equation}
 \Delta F^{prox}_t = \sqrt{(\Delta F^{fc}_t)^2+(\Delta F^{rms}_t)^2+(\Delta F^{var}_t)^2+(\Delta F^z_t)^2}.
 \label{eq:f_er_total}
 \end{equation}
 Similar treatment on the error budget is carried out for the
  estimation of error in the flux from the control sample except that
  $\Delta F^{var}_t$ and $\Delta F^z_t$ are ignored here. The $\Delta
  F^z_t$ does not exists for IGM and $\Delta F^{var}_t$ is negligible
  due to large statistics, having $\sim$ 25 quasars in the control
  sample for each member of the pair in our sample. Therefore, the
  total error on the transmitted flux from the control sample ($\Delta
  F^{\textsc{igm}}_t$) in a given radial distance bin is,
 \begin{equation}
 \Delta F^{\textsc{igm}}_t = \sqrt{(\Delta F^{fc}_t)^2+(\Delta F^{rms}_t)^2}.
 \label{eq:f_er_total}
 \end{equation}
 After taking into account all these uncertainties, we can compare the
 transmitted flux in the longitudinal and transverse directions to that
 in their corresponding control sample, as explained below.

  \begin{figure*}
    \centering
                \includegraphics[height=15cm,width=20cm,trim={4cm 2cm 0cm 0cm}]{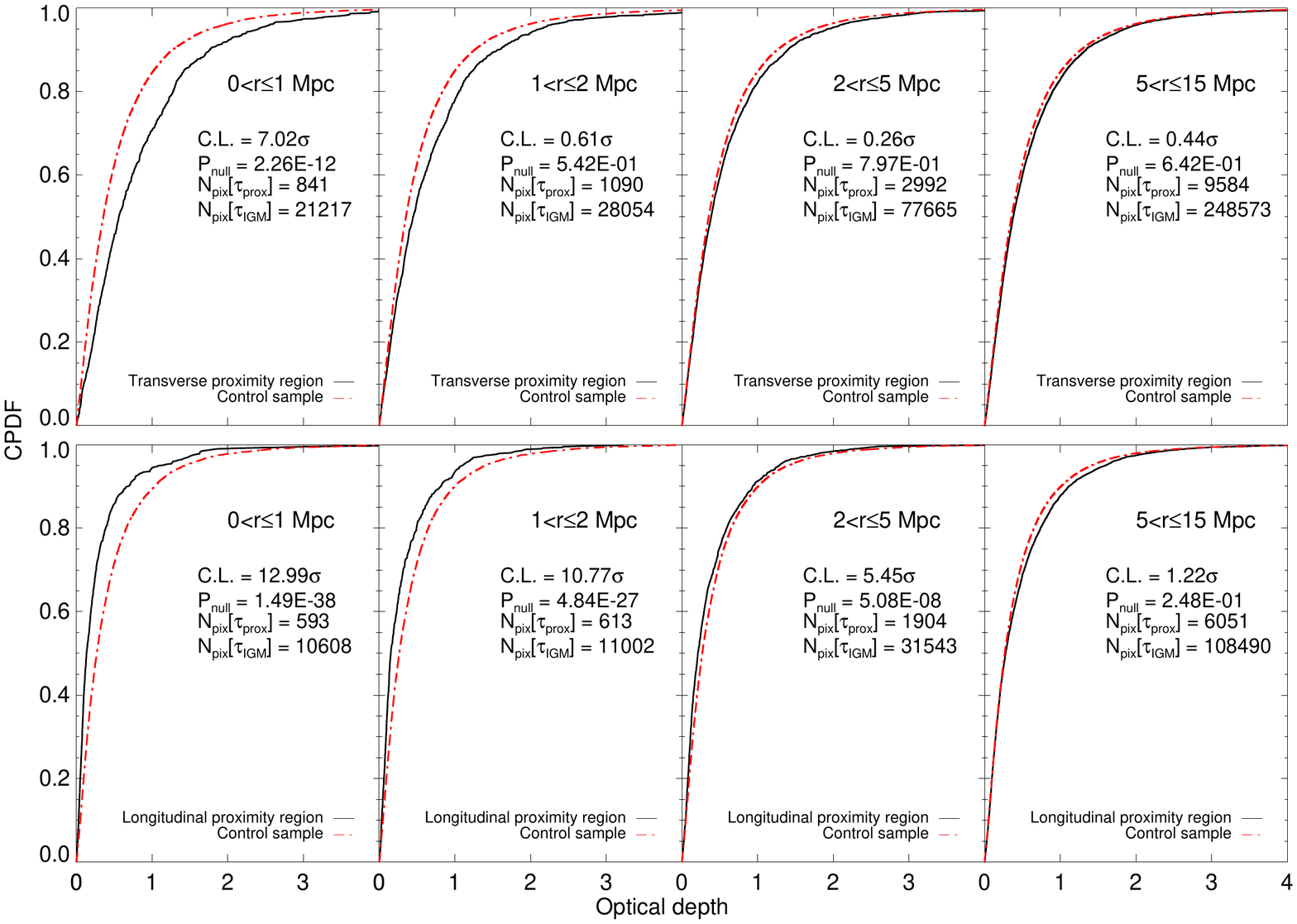}
        \caption {{\it Upper panels: }The cumulative probability
          distribution function (CPDF) of H~{\sc i} \lya optical depth
          in the transverse proximity regions (black-solid) at
          different radial distances from the foreground quasar
          together with the optical depth CPDF from the control sample
          (red-dashed). Labels in each panel give the range of radial
          distance from the foreground quasars used to make the
          subsample for these CPDFs of pixel optical depth (see,
          Sect.~\ref{s:op_dep} for more details). The uncertainty in
          the optical depth at each pixel in the proximity region is
          used to estimate the typical error in the KS-distance
          measurement of KS-test and hence the corresponding
          confidence level (C.L.)  as well as the null probability
          (P$_{null}$) on the measured difference in the CPDFs as
          labeled in each panel (e.g., see Sect.~\ref{s:op_dep}).
            The number of pixels used in the CPDFs of the proximity
            region (N$_{pix}[\tau_{prox}$]) and control sample
            (N$_{pix}[\tau_\textsc{igm}$]) are also mentioned.
          {\it Lower panels: }Same as the top panels but for the
          longitudinal proximity region.}
        \label{Fig:cpdf_tpe}
  \end{figure*}
\subsubsection{Transmitted flux statistics}
\label{s:Analysis1}
 In Fig.~\ref{Fig:rad_flux_red}, we plot median values of the
 normalized transmitted flux, $F_t$ (within 1 \Mpc~radial distance
 bin) at various radial distances from the quasars in both
 longitudinal and transverse directions, towards observer and towards
 background quasar of the foreground quasar. The procedure is similar
 to that given by \citet[][]{Kirkman2008MNRAS.391.1457K} where they
 plot $1-F_t$ in the proximity region of the foreground quasar. \par
 As a consistency check, it can be seen from the
 Fig.~\ref{Fig:rad_flux_red} that the values of median $F_t$ at
 different radial distance bins are almost constant (within their
 uncertainties) for both the control samples used in the LPE and TPE
 analysis.  Furthermore, it can also be noted that $F_t$ measured in
 the proximity region is consistent with that from the control sample
 at large radial distances (as expected) with its median value of
 $\sim$ 0.80. It may be pointed out that in each radial distance bin,
 we have lumped together the pixels over the redshift range of $\sim$
 2.5 to 3.5 due to scatter in the $z_f$ without accounting for the
 existing optical depth redshift evolution. However, we stress that
 our appropriate selection of the redshift matched control samples
 (e.g., see Sect.~\ref{s:control_sample}) take care of such effects
 due to its similar redshift evolution.  \par From
 Fig.~\ref{Fig:rad_flux_red}, we can see the difference in $F_t$ among
 the longitudinal (left panel) and the transverse (right
   panel) directions. It can be seen from the left-hand panel of this
 figure that there is a clear increase of the transmitted flux as we
 go closer to the quasar ($r\leq 4$ \Mpc, towards to observer)
 in the longitudinal direction as compared to its control sample. This
 hints towards the dominance of the quasars ionization, as expected in
 the classical proximity effect, at smaller radial distances. Note,
 \citet[][]{Kirkman2008MNRAS.391.1457K} did not detect such a signal
 of the longitudinal proximity effect. It can be noted from this
   figure that the transmitted flux redward of the \lya emission
   (shown by fainter color) in the first three radial distance bins
   (i.e., $0 \leq r \leq 3$ Mpc) show nominal absorption
   features. Such absorption could arise due to the possible redshift
   uncertainty (even after incorporating the average systematic
   redshift correction, e.g., see Sect.~\ref{s:zq}) and/or the
   possible inflow of the \lya clouds towards the center of quasar's
   halo (resulting in the redshift of such clouds to be more than the
   emission redshift of the foreground quasars). At higher radial bin
   (i.e., $r>3$ Mpc) the transmitted flux is close to unity albeit a
   small systematic decrement of about 2\%, similar to the residual
   absorption found in this region by
   \citet[][]{Kirkman2008MNRAS.391.1457K} where they have ascribed it
   to metal absorption. Nonetheless, we still checked this region
   visually to ensure that continuum is not systematically
   overestimated in individual spectra. \par However, from our
 analysis in the transverse direction, this trend appears to be
 reversed, especially within 2 Mpc radial distance blueward and
 redward of the foreground quasar \lya emission line. This clearly
 suggests the presence of excess H~{\sc i} absorption closer to the
 quasar in the transverse direction likely due to less illumination in
 this direction (e.g., see Sect.~\ref{s:ionization}).  \par
 Additionally, we may also recall that in our emission redshift
 estimation we have added a systematic shift of 229 \kms~\citep[as
   estimated by][]{Shen2016ApJ...831....7S} so it will be interesting
 to see its impact on the above analysis. For this, we have repeated
 our analysis by using emission redshift without applying this
 systematic shift as shown in Fig.~\ref{Fig:rad_flux}. As can be seen
 from this figure that a clear asymmetry in H~{\sc i} absorption is
 evident between the absorption profile blueward and redward of the
 foreground quasar in the transverse direction. Such asymmetry could
 be easily misinterpreted as a consequence of a possible episodic
 lifetime of the quasars \citep[e.g.,
   see][]{Kirkman2008MNRAS.391.1457K,Khrykin2016ApJ...824..133K}.
 Additionally, a consonance may be noted between the systematic shift
 of 229 \kms~in the emission redshift provided by
 \citet[][]{Shen2016ApJ...831....7S} from an independent method to the
 exact symmetric H~{\sc i} absorption profile measured in the
 transverse direction (when correction is applied). In what follows,
 we always use the emission redshifts corrected for the above
 mentioned systematic uncertainty.

   \begin{figure}
    \centering
    \includegraphics[height=7cm,width=7cm]{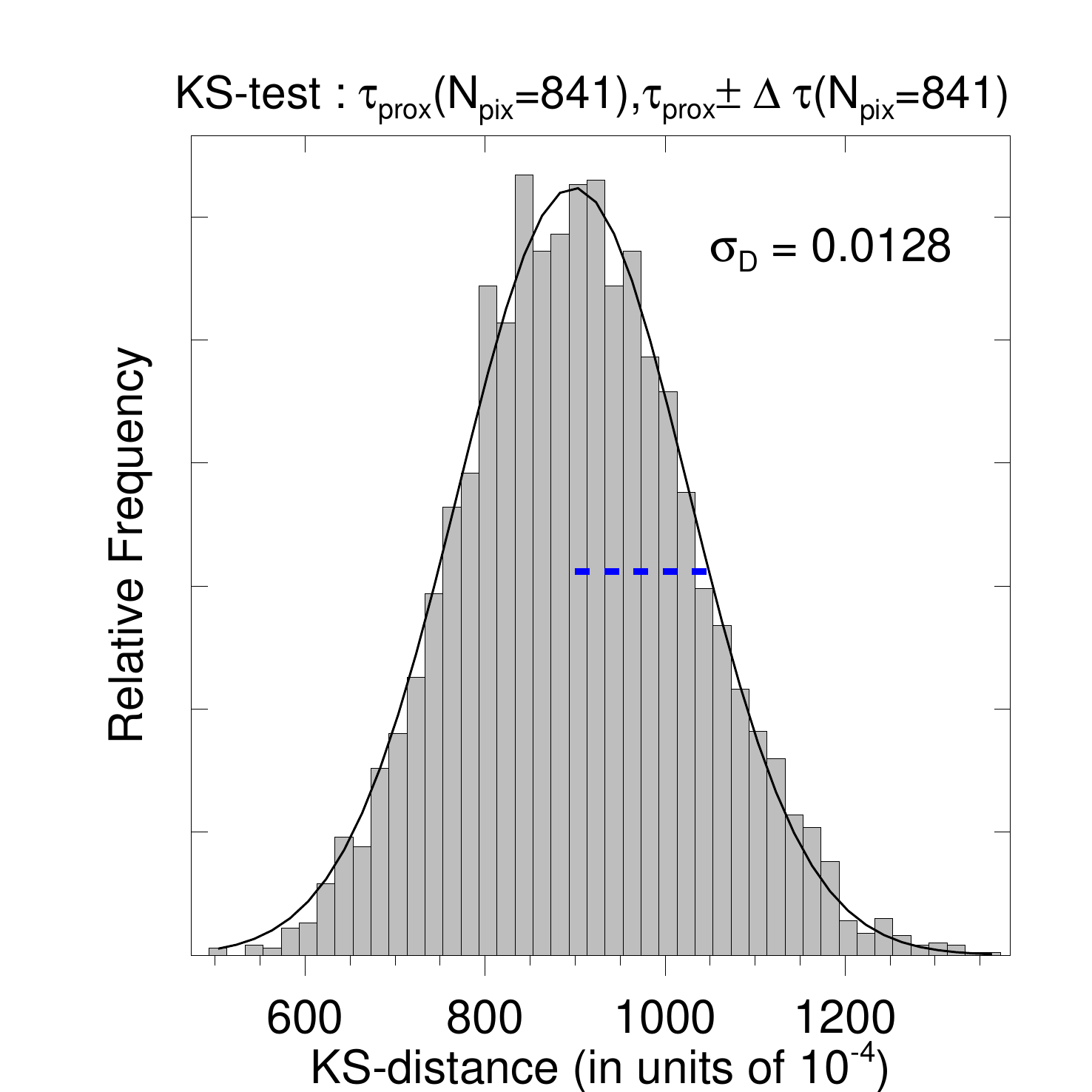}
\includegraphics[height=7cm,width=7cm]{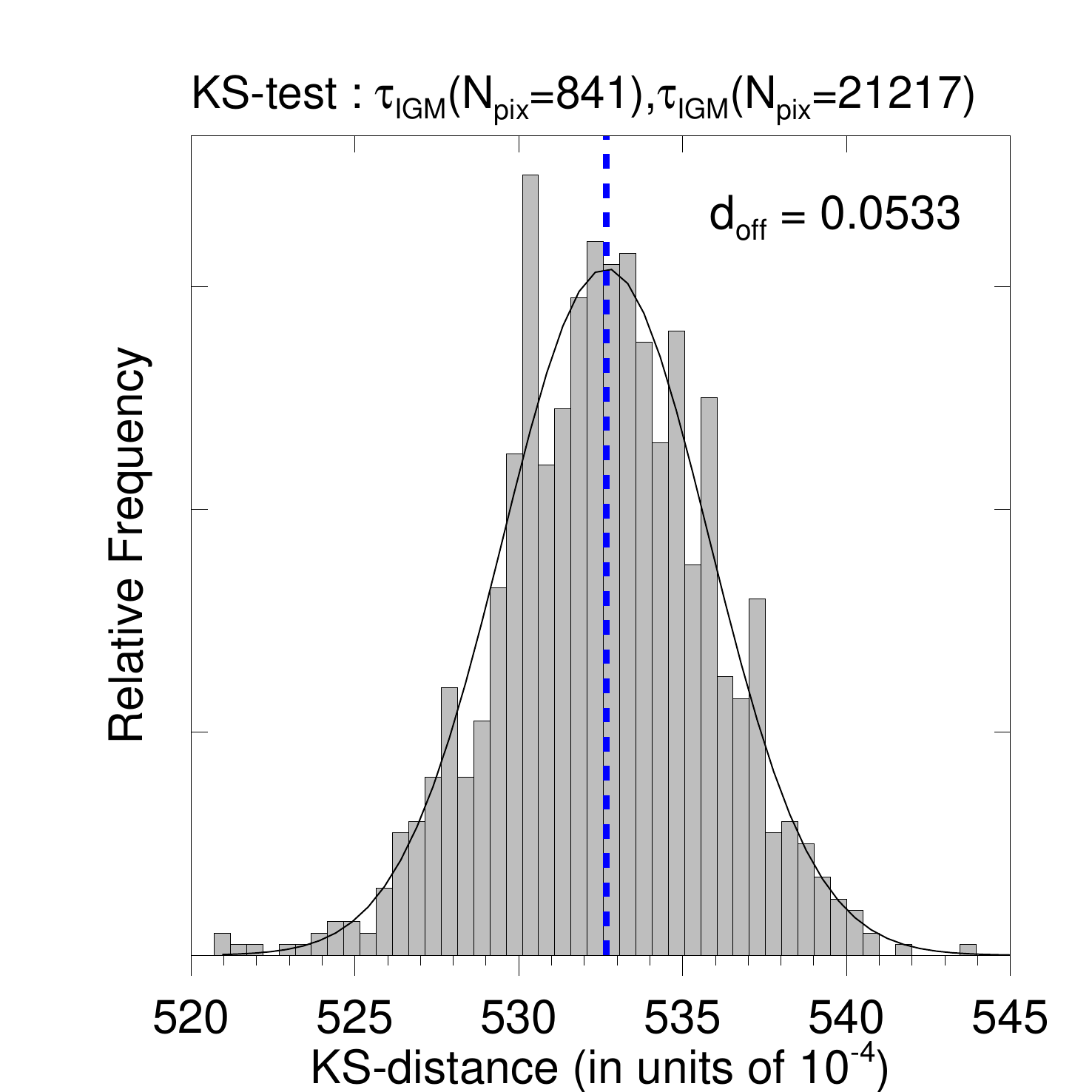}
        \caption {{\it Upper panel:} The plot shows the distribution
          of the KS-distance values obtained by comparing the CPDF of
          the pixel optical depth of the proximity region
          ($\tau_{prox}$) with its 5000 realizations ($\tau_{prox} \pm
          \Delta \tau$), generated assuming the Gaussian distribution
          at each pixel, with a width corresponding to the error bars
          of the pixel optical depth. The plotted distribution of the
          measured KS-distances allows us to compute its standard
          deviation ($\sigma_{\textsc{d}}$). {\it Lower panel:} The
          plot shows the distribution of the KS-distance values
          obtained by comparing the CPDF of the pixel optical depth of
          the parent sample of the IGM ($\tau_\textsc{igm}
          [N_{pix}=21217]$) with its 1000 random subsets
          ($\tau_\textsc{igm} [N_{pix}=841]$). The number of pixels in
          each of the subset is set to be equal to the number of
          pixels available in their corresponding proximity
          sample. The plotted distribution of the measured
          KS-distances allows us to compute its mean value
          (d$_{off}$), used to apply the offset correction in the
          measured KS-distance between the proximity vis-a-vis the IGM
          distribution and hence the confidence level as
          [D$_{ks}$-d$_{off}$]/$\sigma_{\textsc{d}}$.}
        \label{Fig:ks}
  \end{figure}
   
\subsubsection{Pixel optical depth statistics}
\label{s:op_dep}
 We use the values of fluxes used to obtain median fluxes plotted in
 Fig.~\ref{Fig:rad_flux_red}, to compute the optical depth as
 $\tau(\lambda_i)= -ln[F_{t}(\lambda_i)]$ (e.g., see Eq.~\ref{eq:ft})
 for our subsequent pixel optical depth analysis. The pixel optical
 depth we obtain is an integrated value over the pixel width (i.e.,
 $\sim$ 69 \kms~in our SDSS spectra having $R\sim$ 2000). Using these
 pixel optical depth values, we have computed the cumulative
 probability distribution function (CPDF) as shown in
 Fig.~\ref{Fig:cpdf_tpe}. The figure compares the CPDFs of the optical
 depth in the proximity region with that of the control sample within
 the radial distance bins of 0-1, 1-2, 2-5 and 5-15 \Mpc. As can be
 seen from this figure that the form of the optical depth distribution
 in the proximity region is similar to the distribution in the control
 sample. However, its value in the transverse proximity region is
 higher than its control sample (e.g., see upper panel), while the
 trend seems to be reversed in the longitudinal direction (e.g., see
 lower panel), at least up to the radial distance of $\le5$ Mpc from
 the ionizing foreground quasar.\par To quantify the difference
 evident in the CPDFs of the proximity and the corresponding control
 sample, we have used the Kolmogorov-Smirnov (KS) test. This allows us
 to compute the KS-distance (D$_{ks}$) and the probability for the two
 distributions to be the same (P$_{null}$).  However, we note that the
 P$_{null}$ and D$_{ks}$ estimated in the standard KS-test does not
 take into account the uncertainties on the pixel optical depth and
 hence may lead to an overestimation of the significance level of any
 difference among the two CPDFs. The resultant distortion introduced
 by such uncertainties can be significant for the optical depth CPDFs
 of the proximity region due to its small statistics, though any such
 effect will be negligible in the control sample having larger
 statistics (being $\sim$ 25 times that of proximity
 region). Therefore, to be on the conservative side, instead of using
 the standard KS-test, we have estimated the significance of the
 measured D$_{ks}$ and P$_{null}$ by taking into account the
 uncertainties in the optical depth of the proximity region (as also
 used in Fig.~\ref{Fig:rad_flux}), as follows. \par We assume the
 measured value of the optical depth and its uncertainties at a pixel
 as a mean and 1$\sigma$ width of a Gaussian distribution of the
 optical depth at that pixel. Then using this distribution, 5000
 random values of optical depth are generated at each pixel. The
 distribution of KS-distance values obtained by comparing the CPDFs of
 these 5000 realizations with the original CPDF of the proximity
 region is found to be well-fitted with a Gaussian profile. The upper
 panel of Fig.~\ref{Fig:ks} shows an example of one such distribution
 of KS-distance values for a radial distance bin. The 1$\sigma$ width
 of this KS-distance distribution, $\sigma_{\textsc{d}}$, is
 considered as a typical uncertainty in the measured D$_{ks}$. Now, in
 principle, the ratio of the D$_{ks}$ and $\sigma_{\textsc{d}}$ can be
 used to determine the confidence level, however, we noticed that the
 measured mean value of KS-distance has a non-zero offset when we
 compare two distributions with a different number of data points,
 even when they are drawn from the same parent distribution. Such an
 offset can artificially lead to an over-estimation of the confidence
 level based on the ratio of the D$_{ks}$ and
 $\sigma_{\textsc{d}}$. In order to account for this bias, we use the
 control sample and compare it with its own random subsamples. These
 random subsamples are constructed with a constraint that the number
 of pixels in these subsamples should be equal to that in their
 corresponding proximity sample. \par Using the comparison of such
 1000 subsamples with their own parent control sample, we computed the
 mean KS-distance offset (d$_{off}$). An example of one such
 distribution of KS-distance values to compute $d_{off}$ is plotted in
 the lower panel of Fig.~\ref{Fig:ks}. This allows us to compute the
 accurate confidence level (C.L.) as
 [D$_{ks}$-d$_{off}$]/$\sigma_{\textsc{d}}$ and P$_{null}$ as the area
 under the normalized Gaussian curve of standard deviation
 $\sigma_{\textsc{d}}$ beyond $\pm$[D$_{ks}$-d$_{off}$] range, as
 labeled in each panel of the Fig.~\ref{Fig:cpdf_tpe}. It can be noted
 from the figure that in the transverse direction the measured optical
 depth in the proximity region is higher than the control sample in
 0-1 Mpc radial bin at a significance of 7.02$\sigma$. Also, while
 plotting the CPDF for the transverse direction, we have lumped
 together the absorption systems in blueward and redward of the
 foreground quasar \lya emission line. The difference in the
 longitudinal direction is also significant (e.g., see lower panel),
 though have an opposite trend with being smaller in the proximity
 region than in the control sample at 12.99$\sigma$ and 10.77$\sigma$
 in the radial bins of 0-1 and 1-2 Mpc respectively. \par This
 dissimilarity of the CPDFs of the optical depth in the transverse and the
 longitudinal directions indicates that the observed optical depth
 distribution around the quasar may be anisotropic. This is consistent
 with the inference drawn based on the transmitted flux statistics
 (e.g., see Fig.~\ref{Fig:rad_flux_red}). This anisotropic
 distribution could result from an anisotropic radiation field from the
 quasars or anisotropic matter distribution around them. However,
 before interpreting the difference seen in the CPDFs, the value of
 optical depth at a given pixel should be corrected for the effect of
 quasar's ionizing radiation as we describe in the next subsection.

    \subsection{Ionization and overdensity effects in the proximity region}
    \label{s:ionization}
 From our analysis of the transmitted flux (e.g., see
 Fig.~\ref{Fig:rad_flux_red}) and the pixel optical depth
 distributions (e.g., see Fig.~\ref{Fig:cpdf_tpe}) it is evident that
 these quantities distribute differently along the transverse and the
 longitudinal directions. The role of anisotropic matter distribution
 causing this observed difference seems to be unlikely because we have
 used a large sample of quasars that can lead to an average density
 profile being isotropic around the quasar. Based on this, we can
 ascribe the differences we find to anisotropy in the quasar
 ionization.  To quantify this anisotropic distribution, we will first
 estimate the quasar's ionization based on the observed flux in the
 longitudinal direction to get the ionization corrected pixel optical
 depth distribution. This will enable us to estimate the ionization
 corrected average excess overdensity profile in the longitudinal
 direction. Then, we will be keeping the fraction of quasar's
 illumination in the transverse direction compared to the longitudinal
 direction as a free parameter and constrain its best fit value by
 ensuring a statistical match of the ionization corrected average
 excess overdensity profile in these two directions.  \par To estimate the
 quasar's ionization, we note that its extent on the observed
 transmitted flux and pixel optical depth will also be influenced by
 the strength of the UVB radiation, along with the amount of
 clustering of matter around quasars expressed in terms of overdensity
 at a given radial distance bin from the quasar. To lift this
 degeneracy, we estimated the excess ionization by the quasar relative
 to the UVB radiation at each radial distance, $r$, from the quasar by
 computing a scaling factor of $[1+\omega_r]$ defined as
 \begin{equation}
 \frac{\Gamma_{\textsc{uvb}}(z_{a})+\Gamma_{q}(r,z_{a})}{\Gamma_{\textsc{uvb}}(z_{a})}
 \equiv 1+\omega_r
 \label{eqn:omega}
\end{equation}
 where, $\Gamma_{\textsc{uvb}}(z_{a})$ and $\Gamma_{q}(r,z_{a})$ are the
 H~{\sc i} photoionization rates at the absorbing cloud at redshift
 $z_{a}$ contributed by the UVB and quasar respectively. The $\Gamma_{\textsc{uvb}}$ is
 defined by
\begin{equation}
  \Gamma_{\textsc{uvb}}(z_{a})=\int^\infty_{\nu_{912}}
  \frac{ 4\pi J_\nu(z_{a})}{h\nu}\sigma_{HI}d\nu
  \label{eqn:igm_photo}
\end{equation}
where, $J_{\nu}$ is the average specific intensity of UVB (in units of
erg~cm$^{-2}$ s$^{-1}$ Hz$^{-1}$ sr$^{-1}$), $\nu_{912}$ is the frequency
corresponding to 912 \AA~(i.e., H~{\sc i} ionizing energy) and
$\sigma_{HI}$ is the photoionization cross-section given by
  \begin{equation}
\label{eq:photoionization cross section}
\sigma_{HI} = \sigma_{0}
\left(
\frac{\nu}
{\nu_{912}}
\right)^{-3}
  \end{equation}
 with $\sigma_{0}=6.30 \times 10^{-18}$ cm$^{2}$
 \citep[e.g.,][]{Osterbrock2006agna.book.....O}. We used the value of
 $\Gamma_{\textsc{uvb}}(z)$ given by \citet[][]{Khaire2015MNRAS.451L..30K}
 based on their updated estimation of $J_{\nu}$ using comoving
 specific galaxy and quasar emissivities at different redshifts. The
 $\Gamma_{q}$ is given by,
\begin{equation}
\Gamma_{q}(r,z_{a})=\int^\infty_{\nu_{912}}\frac{L_\nu}{4\pi
  r^2}\frac{\sigma_{HI}}{h\nu}d\nu
\label{eqn:gammaqso_int}
\end{equation}
 where, $r$ is the distance of the absorbing cloud at $z_{a}$ from
 the foreground quasar at $z_f$ (e.g., see Eq.~\ref{eq:prop_r}) with
 luminosity $L_\nu$ (in units of erg~s$^{-1}$ Hz$^{-1}$) calculated by
 Eq.~\ref{eq:lnu_fnu} and Eq.~\ref{eq:power}.
 Integrating, the above equation results in,
\begin{equation}
\Gamma_{q}(r,z_{a}) \\ = \frac{ \sigma_{0}~L_{\nu_{1325}}}{4\pi r^2 h
  (3+\alpha_{uv})}\left(\frac{\nu_{1300}}{\nu_{1325}}\right)^{-\alpha}\left(\frac{\nu_{912}}{\nu_{1300}}\right)^{-\alpha_{uv}}
\label{eqn:gamma_q}
\end{equation}
where, $h$ is the Planck's constant. With our estimation of
$\Gamma_{q}(r,z_{a})$ and $\Gamma_{\textsc{uvb}}(z_{a})$, we can compute $[1+\omega_r]$
scaling factor to compensate for the decrement of the optical depth due to
the excess ionization \citep[e.g., see][]{Rollinde2005MNRAS.361.1015R,
  Faucher2008ApJ...673...39F} as,
\begin{equation}
  \tau_{off} = [1+\omega_r]\times \tau_{prox}
  \label{eqn:off}
\end{equation}
where, $\tau_{off}$ is the optical depth that would be obtained if the
quasar was turned off and $\tau_{prox}$ is the measured optical depth
in the presence of the quasar. However, such [$1+\omega_r$] scaling
for $\tau_{prox}$ is valid for high-resolution spectra and may not be
valid in the case of the SDSS spectra where the measured optical depth
does not necessarily follow the column density due to poor spectral
resolution \citep[e.g., see][]{Lee2012AJ....143...51L}.
  Therefore, it is essential to carry out a quantitative analysis on
  the validity of such scaling at low/moderate resolution using
  simulated spectra. For this purpose, we generated simulated spectra
based on numerical simulation to quantify a scaling relation of the
form as,
\begin{equation}
  \tau_{off} = [1+\omega^{ef}_r(\tau,\omega_r)]\times \tau_{prox}.
  \label{eqn:off}
\end{equation}
 Here, $[1+\omega^{ef}_r(\tau,\omega_r)]$ is an effective optimized
  scaling for low/moderate resolution spectra instead of the
  [$1+\omega_r$] scaling such that,
 \begin{equation}
  \bigg[\frac{\Gamma_{\textsc{uvb}}(z_{a})+\Gamma_{q}(r,z_{a})}{\Gamma_{\textsc{uvb}}(z_{a})}\bigg]^{ef} \equiv 1+\omega^{ef}_r(\tau,\omega_r) 
   \label{eqn:fw_eff}
 \end{equation}

 \par In addition, quasars may also reside in an overdense region \citep[e.g.,
  see][]{Rollinde2005MNRAS.361.1015R,Kirkman2008MNRAS.391.1457K,Finley2014A&A...572A..31F,Adams2015MNRAS.448.1335A,Lau2018ApJ...857..126L}.
 Therefore, contrary to the
expected decrease in the optical depth due to extra ionization by the
quasar's radiation this overdensity will lead to an increase of the
observed effective optical depth in the proximity region as described
below. \par The photoionization equilibrium, for a highly ionized
optically thin gas is given by,
\begin{equation} 
  n_{\textsc{Hi}}\Gamma_{\textsc{Hi}} = R(T[\Delta])~n_en_{\textsc{Hii}} \approx Rn^2_{\textsc{Hii}}
\label{eq:photo_equil}
\end{equation}
where, $R(T[\Delta]) = 4.2\times 10^{-13}(T(\Delta)/10^4K)^{-0.7}$
cm$^3$ s$^{-1}$ is the recombination rate
\citep[e.g.,][]{Hui1997ApJ...486..599H} and
$T(\Delta)=T_0\Delta^{\gamma-1}$ is the temperature of the gas having an
overdensity $\Delta=\rho(r)/\rhobar$~with an exponent $\gamma$.  Here,
we have neglected the helium contribution to the free electron density
which typically amounts to an error of 8\% on $n_e$ as estimated by
\citet[][]{Faucher2008ApJ...673...39F} for singly ionized helium. \par
The density of ionized hydrogen ($n_{\textsc{Hii}}$) will be the
product of the total hydrogen number density ($n_H$) and the fraction
of ionized hydrogen ($X_{\textsc{Hii}}$) i.e., $n_{\textsc{Hii}} =
n_{H}X_{\textsc{Hii}}$ with $n_H = \bar{n}_H \Delta$. As we know,
$\tau_{\textsc{Hi}} \propto n_{\textsc{Hi}}$ which for a given
$\Gamma_{\textsc{Hi}}$ will be proportional to the product of
$R(T[\Delta])$ and $\Delta^2$ resulting in
\begin{equation}
  \tau \propto n_{\textsc{Hi}} \propto T^{-0.7}\Delta^2 \propto \Delta^{2-0.7(\gamma-1)}.
\end{equation}
 The combined effect of such a density enhancement and extra
ionizing photons around quasars (discussed above, e.g., see
Eq.~\ref{eqn:off}), is to shift the observed optical depth in the
proximity region $\tau_{prox}$. Following approach similar to
\citet[][]{Rollinde2005MNRAS.361.1015R} and using Eq.~\ref{eqn:off}, we
quantify this shift for the measured optical depth in our SDSS spectra
as,
\begin{equation}
        \tau_{\textsc{igm}} =
        \tau_{prox}\,\frac{[1+\omega^{ef}_r(\tau,\omega_r)]}{\left(\rho(r)/\rhobar\right)^{2-0.7(\gamma-1)}}.
        \label{eqn:shiftprox}
\end{equation}
 Therefore, the ratio of $\tau_{prox} \times
 [1+\omega^{ef}_r(\tau,\omega_r)]/ \tau_{\textsc{igm}}$ in various
 radial bins will allow us to estimate an average excess overdensity
 profile around the quasars in the longitudinal direction (e.g., see
 Eq.~\ref{eqn:shiftprox}). The corresponding density profile for the
 transverse direction will be given by $\tau_{prox} \times
 [1+\omega^{ef}_r(\tau,\eta\omega_r)]/ \tau_{\textsc{igm}}$ where,
 $\eta$ is the fraction of the illumination in the transverse
 direction as compared to the longitudinal direction (with maximum
 range of $0 \leq \eta \leq 1$). The best fit value of $\eta$ is
 obtained by $\chi^2$ minimization between the average excess
 overdensity profile in the longitudinal direction and transverse
 direction as discussed in subsequent subsections (e.g., see
 Sect.~\ref{s:over}). Here, we may recall that the observed optical
 depth in the transverse direction is already higher than in the
 longitudinal direction, therefore, $\eta>1$ will not be consistent
 with our assumption of the spherically symmetric distribution of
 matter around the quasar.  Before that we first quantify in the next
 subsection, the scaling factor [$1+\omega^{ef}_r(\tau,\omega_r)$]
 that is required to estimate the appropriate ionization correction
 for scaling the optical depth measured in the low/moderate resolution
 SDSS spectra as a function of $\omega_r$ and $\tau_{prox}$.
    \begin{figure}	
  \centering
  \includegraphics[height=10cm,width=7.5cm,trim={1cm 0cm 1cm 0cm}]{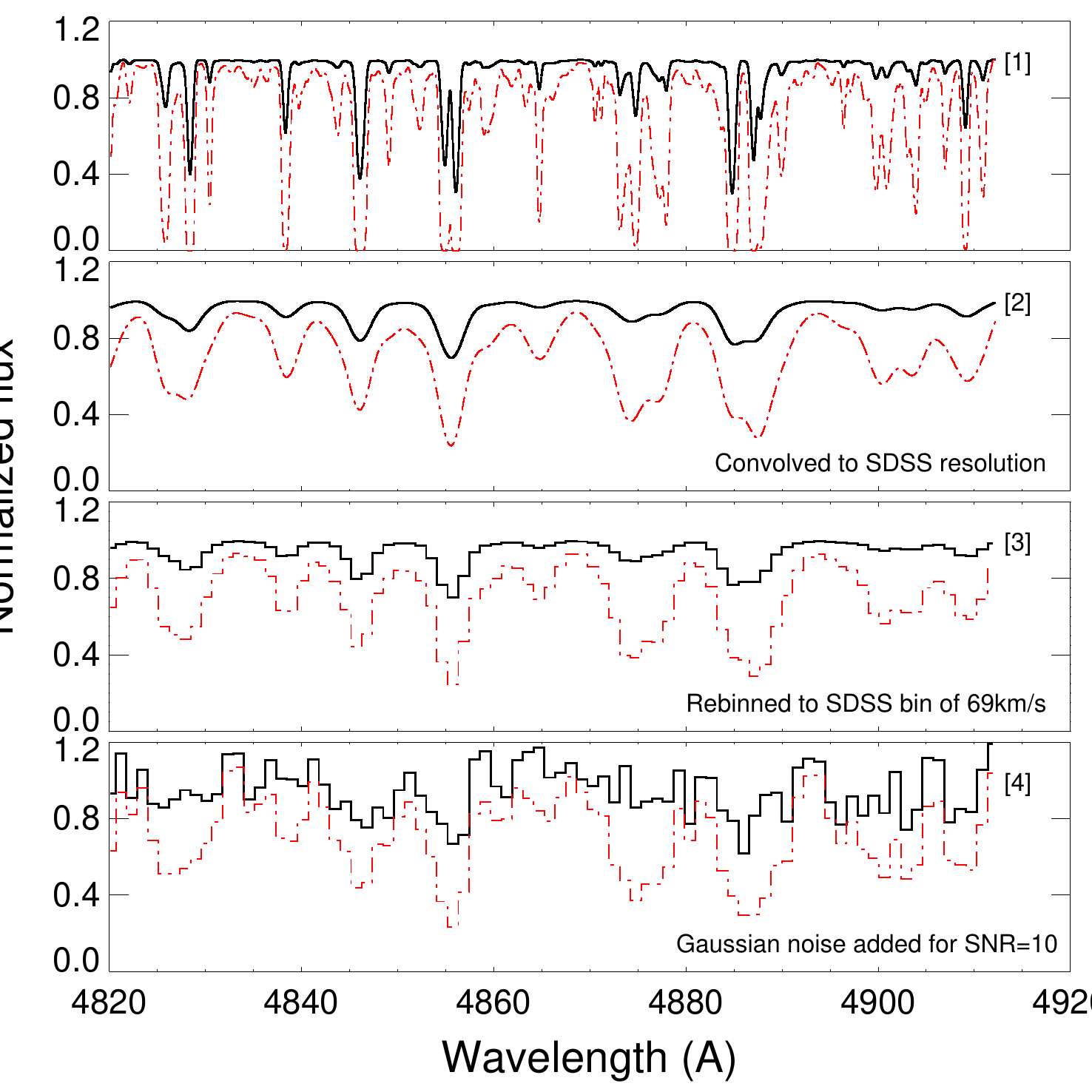}
  \caption {{\it Panel [1]: } Transmitted flux along a line of sight
    in our simulation box with the gas being ionized by the UVB (red
    dotted) and excess radiation from a nearby ionizing source (i.e.,
    UVB and quasar, black solid). The additional quasar ionization
    effect is mimicked by scaling down the optical depth in each pixel
    by a factor $[1+\omega_r]\approx 11$.  {\it Panel [2]: } This
    panel shows the spectra displayed in the panel [1] after its
    convolution with a Gaussian function having FWHM corresponding to
    the SDSS resolution of $R\sim 2000$. {\it Panel [3]:} This panel
    shows re-binning of the spectra in the panel [2] to a SDSS pixel
    width of 69 \kms. {\it Panel [4]:} This panel shows the spectra of
    the panel [3] after adding a random Gaussian noise corresponding
    to a typical SNR ($\approx 10$) of our real SDSS spectra. }
      \label{Fig:simu_sightline}
    \end{figure}


\subsubsection{Appropriate ionization corrections for moderate
    resolution spectra}
\label{s:simulation}

  We have used the hydrodynamical simulations of IGM, as discussed in
  details by \citet[][]{Gaikwad2018MNRAS.474.2233G}. In brief, the
  simulated spectra are obtained by shooting sightlines through the
  simulated box of size $\sim$ 14 \Mpc~(comoving) having $2 \times 512^3$
  number of particles. This small box size might amount to assume
    no spatial correlations of the optical depth between the various
    radial distance bins. However, its impact may not be significant
    as our aim here is to validate the ionization scaling only for an
    effective pixel optical depth value instead for an optical depth
    distribution. We generated 200 realizations of \lya optical depth
  ($\tau_{true}$) as a function of wavelength at a median redshift of
  2.5 and 3.0 (e.g., see panel [1] of Fig.~\ref{Fig:simu_sightline},
  red-dotted line). Using these simulated spectra, we estimated the
  optimal ionization scaling (e.g., see Eq.~\ref{eqn:off}) that is
  appropriate for our low/moderate resolution SDSS spectra as follows:
  \ben
\item We have used the simulated pixel optical depth ($\tau_{true}$,
  e.g., see panel [1] of Fig.~\ref{Fig:simu_sightline}, red-dotted line) to
  generate the mock spectra at the SDSS resolution ($\tau_{\textsc{igm}}$). For
  this, we first convolve the simulated spectrum (i.e.,
  $e^{-[\tau_{true}]}$) with a Gaussian function having a FWHM
  corresponding to the SDSS resolution of $R \sim 2000$ (e.g., see panel [2]
  of Fig.~\ref{Fig:simu_sightline}, red-dotted line). Secondly, we re-bin
  this convolved spectrum at 69 \kms~interval corresponding to the
  SDSS pixel width (e.g., see panel [3] of
  Fig.~\ref{Fig:simu_sightline}, red-dotted line). Thirdly, to mimic the
  noise of the real spectrum, we added a random Gaussian noise (e.g.,
  see panel [4] of Fig.~\ref{Fig:simu_sightline}, red-dotted line) with
    a mean of zero and a standard deviation of 1/SNR.
\item To generate the optical depth distribution as observed after
  excess ionization due to the quasar observed in a low resolution
    spectrum ($\tau_{prox}$), we follow the same steps as listed
  above (i.e., point [1]) except that instead of using $\tau_{true}$,
  we have used $\tau_{true}/[1+\omega_r]$ for a given input $\omega_r$
  (e.g., see Eq.~\ref{eqn:omega}) as shown by black-solid line in
  panels [1]-[4] of Fig.~\ref{Fig:simu_sightline}.
   \item We noticed that in order to recover the $\tau_{\textsc{igm}}$
     from its corresponding ionized pixels $\tau_{prox}$ (for a given
     $\omega_r$) at the SDSS resolution, the required ionization
     correction i.e., $\omega^{ef}_r(\tau)$ (e.g., see
     Eq.~\ref{eqn:fw_eff}), will strongly depend on the value of
     measured pixel optical depth. Typically, 95\% of our measured
     optical depth in the real spectra lie within the range of 0.01 to
     3 (e.g., see Fig.~\ref{Fig:cpdf_tpe}). Therefore, we carry out
     our analysis by binning the optical depth obtained after
     convolving the simulated spectra at the SDSS resolution. The
     binning is done with non-uniform bins of 0.0-0.01, 0.01-0.02,
     0.02-0.05, 0.05-0.10, 0.10-0.20, 0.20-0.30, 0.30-0.50, 0.50-1.00
     and last bin with $\tau > 1.0$ (bin sizes are optimized to ensure
     reasonable statistics in each bin). As a result, now within each
     optical depth bin, we have a distribution of $\tau_{prox}$ and
     its exact counterpart pixels of $\tau_{\textsc{igm}}$ (i.e.,
     without ionization, see point [1] above) for a typical SDSS
     resolution, by using all the 200 simulated sightlines for both
     the redshifts (i.e., 2.5 and 3.0).

     \begin{figure}	
       \centering
       \includegraphics[height=9cm,width=9.0cm,trim=0 0 0 0]{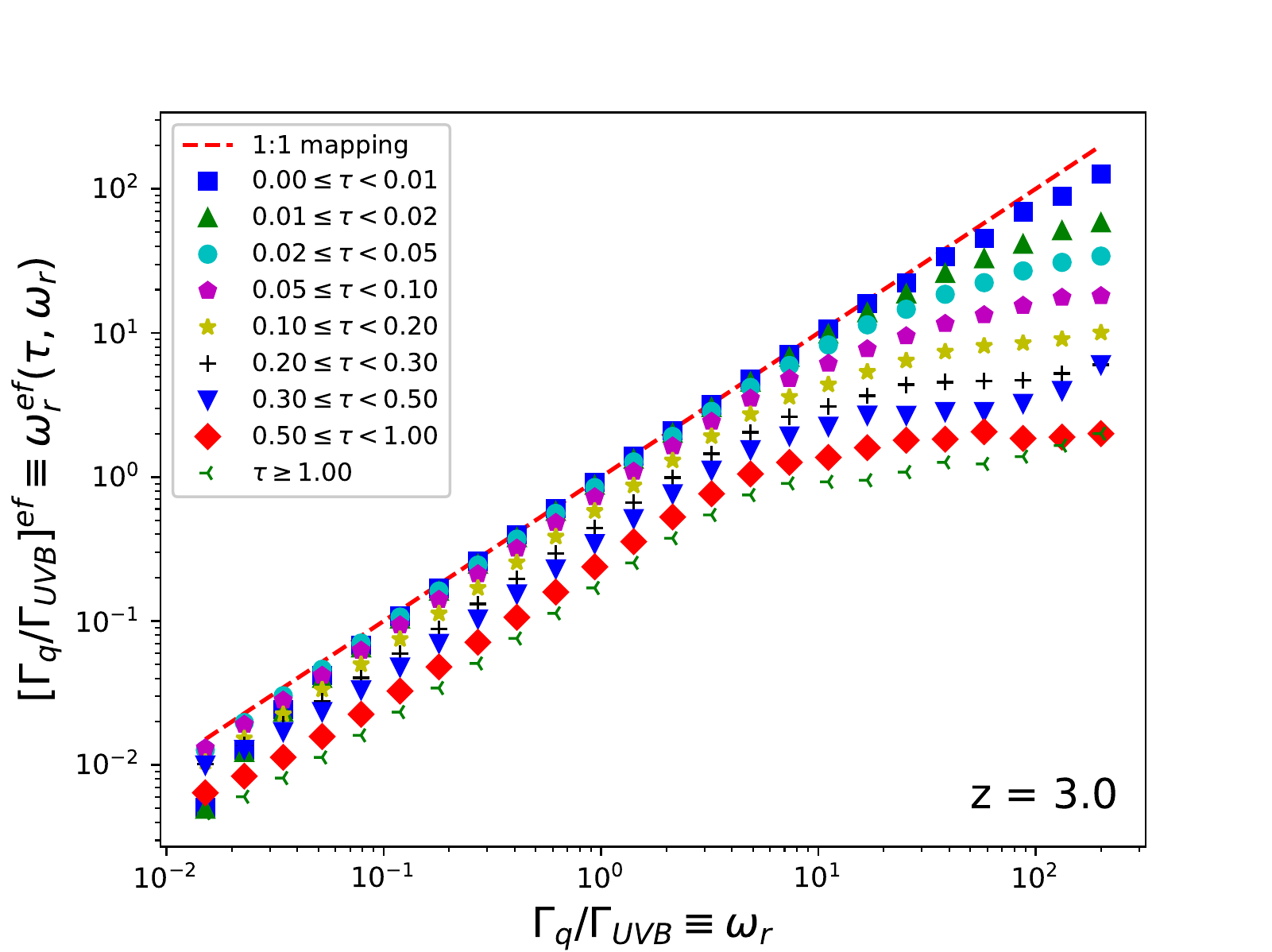}
       \caption {The plot shows the departure of the effective
         ionization correction parameter,
         $[\Gamma_q/\Gamma_{\textsc{uvb}}]^{ef} (\equiv
         \omega^{ef}_r(\tau,\omega_r)$), from its theoretical value of
         $\Gamma_q/\Gamma_{\textsc{uvb}} (\equiv \omega_r$) in various
         pixel optical depth bins (listed in inset) obtained from
         spectra at low SDSS resolution ($R\sim$ 2000). Here, the
         $\omega_r$ is used to ionize the simulated IGM spectra as
         $\tau_{true}/[1+\omega_r]$. The $\tau_{true}$ and
         $\tau_{true}/[1+\omega_r]$ are convolved with Gaussian kernel
         corresponding to the SDSS resolution to obtain
         $\tau_{\textsc{igm}}$ and $\tau_{prox}$ respectively (e.g.,
         see points shown in panel [5] of
         Fig.~\ref{Fig:simu_sightline}). The best fit is computed such
         that KS-test probability is maximum for $\tau_{prox}\times[1+
           \omega^{ef}_r(\tau,\omega_r)]$ and $\tau_{\textsc{igm}}$ to belong
         to a similar distribution at each optical depth bin (binned
         in $\tau_{prox}$).}
         \label{Fig:w_fw}
     \end{figure}

     \item For the actual input ionization correction of
       [1+$\omega_r$], we consider the best fit value of output
       optimal ionization correction to be
       $[1+\omega^{ef}_r(\tau,\omega_r)]$ in a given optical depth
       bin. For this, we use KS-test and compared the distribution of
       $\tau_{\textsc{igm}}$, with various distributions of
       $\tau_{prox}\times[1+\omega^{ef}_r(\tau,\omega_r)]$, by varying
       the trial values of $\omega^{ef}_r(\tau)$ over a range of 1\%
       to 200\% of input $\omega_r$. The value of
       $[1+\omega^{ef}_r(\tau,\omega_r)]$ resulting in maximal of
       $P_{null}$ by KS-test is selected as the best fit optimal
       value. We note that this optimized model value being for the
       mean value of the optical depth bin could also have an
       associated uncertainty.  However, due to very small bin size of
       optical depth used here, this uncertainty is found to be
       negligible in comparison to the errors propagated to it based
       on the uncertainties in estimating the value of the pixel
       optical depth and its redshift as discussed in
       Sect.~\ref{s:over}. \par The above procedure was repeated for
       various $\omega_r$ values ranging from 0.01 to 200 as shown in
       Fig.~\ref{Fig:w_fw} for all the optical depth bins mentioned
       above. The plot is shown only for simulations at $z\sim 3.0$ as
       we found no significant evolution with the redshift. From this
       figure, it is evident that the required ionization correction
       departs towards a lower value compared to the theoretical
       equality line (shown as a red-dashed line in
       Fig.~\ref{Fig:w_fw}) with the increase in the pixel optical
       depth value. This is not surprising as higher optical depth
       pixels generally corresponds to the core of saturated lines
       where the optical depth reduction due to the ionization will be
       negligible. Therefore, the required ionization correction (to
       recover the unionized optical depth) will also be small. For
       instance, it can be noted from Fig.~\ref{Fig:w_fw} that for a
       optical depth of $\sim$ 0.15 and for the typical range of
       0.1-100 in $\omega_r$, the corresponding optimized
       $f(\omega_r)$ varies over a small range of 0.1-5. This clearly
       demonstrates the need for optimizing the ionization correction
       (which is dependent on the pixel optical depth) for the
       analysis of the pixel optical depth based on the low-resolution
       spectrum (as in our case).  \een

 \subsubsection{Pixel optical depth analysis using appropriate ionization correction}
 \label{s:over}

\begin{figure*}   
  \centering
  \includegraphics[height=8cm,width=8cm,trim={0cm 0cm 0cm
      0cm}]{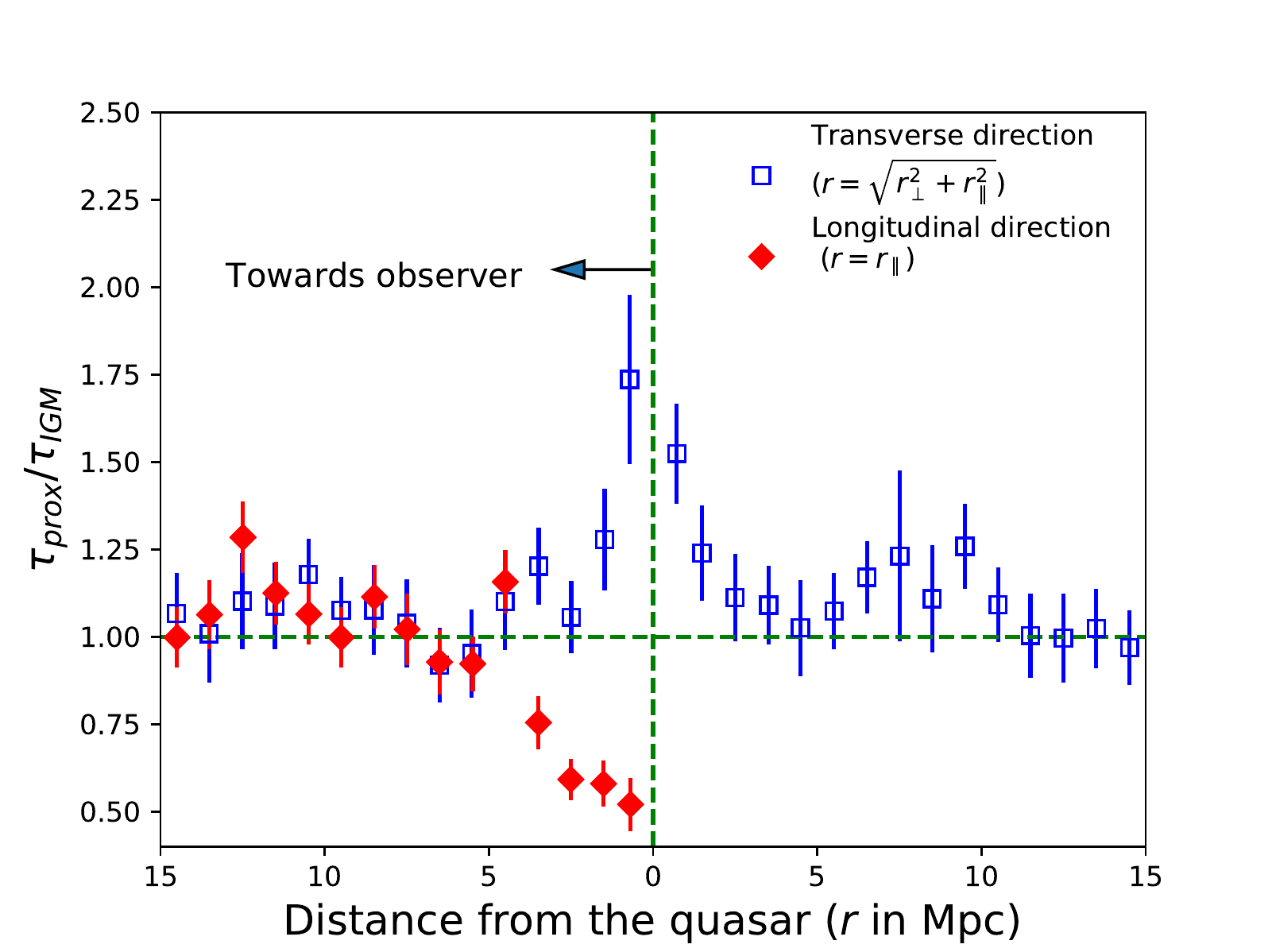}
  \includegraphics[height=8cm,width=8cm,trim={0cm 0cm 0cm
      0cm}]{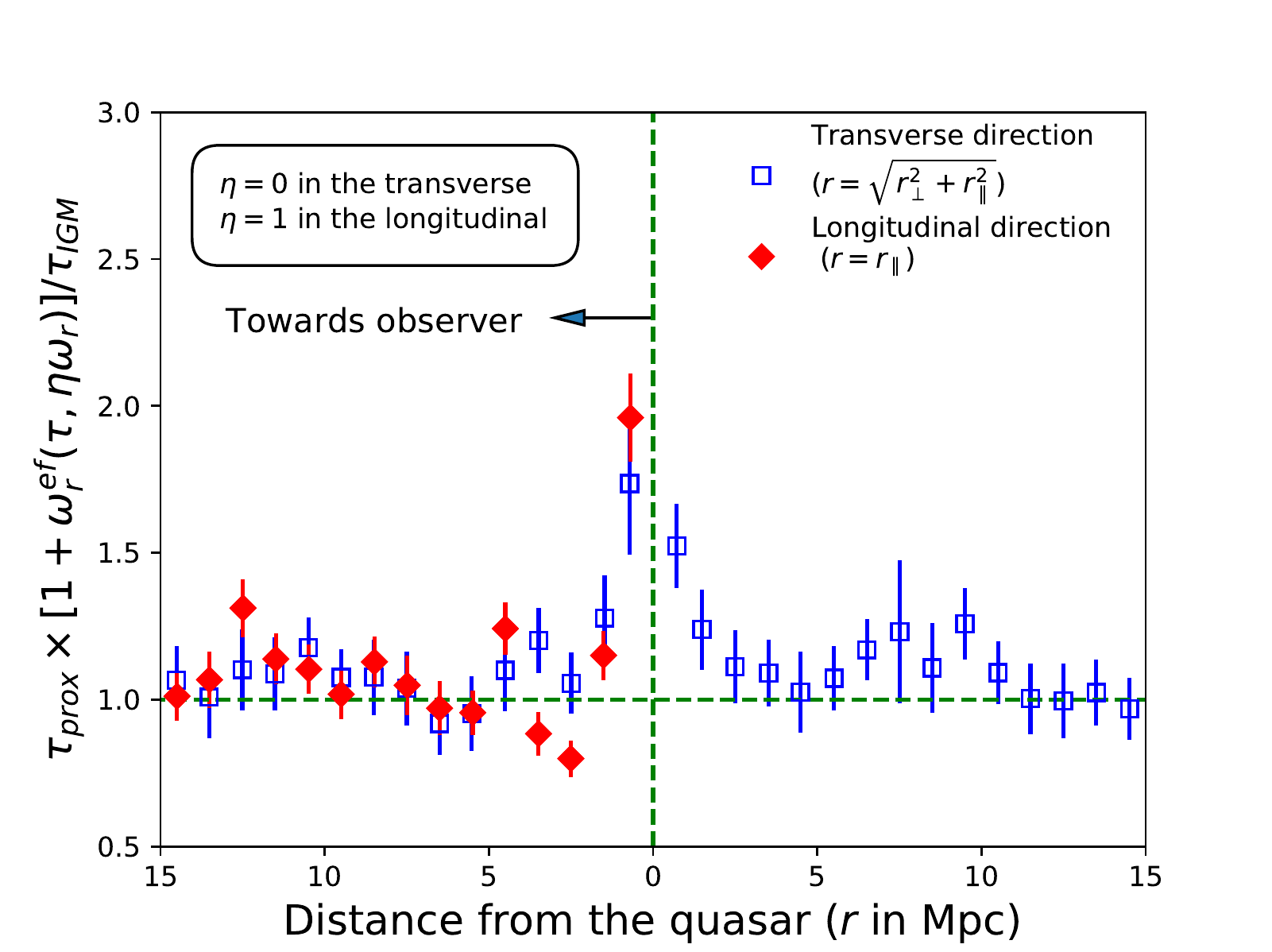}
  \caption {{\it Left panel: }Ratio of the median optical depth value
    (within 1 \Mpc~bin) of proximity region ($\tau_{prox}$) to its
    median value in IGM ($\tau_{\textsc{igm}}$), showing a clear
    discrepancy at small radius between the measurements along the
    longitudinal (red, diamond) and the transverse (blue, square)
    directions. The resultant error bars consist of all the possible
    sources of errors such as flux error from photon counting
    statistics, error due to continuum placement uncertainties, r.m.s
    statistical error, sightline-to-sightline variance and emission
    redshift measurement error within the 1 \Mpc~radial distance bin
    as also used in Fig.~\ref{Fig:rad_flux} (e.g.,
    Sect.~\ref{s:Uncertainties}).  {\it Right panel: }Same
    as left panel but with optical depth scaled by
    [$1+\omega^{ef}_r(\tau,\omega_r)$] for pixels belonging to the
    longitudinal proximity region (i.e., $median(\tau_{prox}\times
    [1+\omega^{ef}_r(\tau,\omega_r)])/median[\tau_{\textsc{igm}}$]).
    Additionally, we also considered here the uncertainty in the
    scaled optical depth due to the uncertainties in estimating
    $\omega^{ef}_r(\tau,\omega_r)$ (e.g., see
    Sect.~\ref{s:simulation}).}
      \label{Fig:omega_crct_overdensity}
\end{figure*}

     In the left panel of Fig.~\ref{Fig:omega_crct_overdensity}, we
     plot the ratio of the median value of the optical depth in the
     proximity region to that of the corresponding IGM,
     $median[\tau_{prox}]/median[\tau_{\textsc{igm}}]$, at various
     radial distance bins of size 1 Mpc. The plot echoes the results
     based on the transmitted flux analysis (e.g., see
     Fig.~\ref{Fig:rad_flux_red}) revealing a clear discrepancy at $r
     \leq 4$ Mpc, between the measurements along the longitudinal and
     transverse directions.  We may recall that the shown error bars
     consist of all the possible sources of errors such as flux error
     from photon counting statistics, error due to continuum placement
     uncertainties, emission redshift measurement error,
     sightline-to-sightline variance and r.m.s statistical error
     within 1 \Mpc~radial distance bin as also used in
     Fig.~\ref{Fig:rad_flux_red} (e.g.,
     Sect.~\ref{s:Uncertainties}). In this comparison, we have
     excluded the radial distances below 0.4 Mpc due to the lack of
     sufficient pixels within such separations along the transverse
     direction, where, a quasar pair can only contribute for pixels
     with $r>r_\perp$. \par Based on the observed luminosity in the
     longitudinal direction, we have estimated the ionization
     corrected $median[\tau_{prox}]/median[\tau_{\textsc{igm}}]$
     (refer to as an ``average excess overdensity profile''), which is
     shown in the right panel of Fig.~\ref{Fig:omega_crct_overdensity}
     (red, diamond). For this, we consider a pixel with the measured
     optical depth value $\tau_i$ at an absorption redshift of
     $z^i_{a}$. Firstly, based on the $\tau_i$ value, we choose the
     corresponding $\omega_r$ versus $\omega^{ef}_r(\tau,\omega_r)$
     curve from those shown in Fig.~\ref{Fig:w_fw}. Then the redshift
     difference between the pixel and the quasar allows us to compute
     the $\omega_r^{i}$ (using the measured luminosity e.g., see
     Eq.~\ref{eqn:omega}). We then estimate
     $\omega^{ef}_r(\tau,\omega_r^{i})$ using the cubic spline
     interpolation of the $\omega_r$ versus
     $\omega^{ef}_r(\tau,\omega_r)$ curve (i.e., from
     Fig.~\ref{Fig:w_fw}). This optimal ionization correction (e.g.,
     see Eq.~\ref{eqn:off}) allows us to estimate the ionization
     corrected optical depth as $\tau_{off} \equiv \tau_{prox}\times
     [1+\omega^{ef}_r(\tau,\omega_r)]$ (also referred to as scaled
     optical depth). Here, the error bar for the $\tau_{off}$ from the
     $\tau_{prox}$ are also propagated appropriately.  \par
     Additionally, we also considered the possible uncertainty in the
     $\tau_{off}$ due to $\omega^{ef}_r(\tau,\omega_r)$
     uncertainty. This error is (i) due to the uncertainty in the
     pixel optical depth ($\Delta \tau$) because of the strong
     dependence of $\omega^{ef}_r(\tau,\omega_r)$ on the optical depth
     value and (ii) due to the uncertainty in $\omega_r$ propagated
     from the uncertainty in emission redshift measurements (reflected
     as $\Delta r$ uncertainty in the distance calculation). The
     former is estimated as;
\begin{equation}
  |\omega^{ef}_r(\tau+\Delta \tau)-\omega^{ef}_r(\tau-\Delta \tau)|/2
    \label{eq:wef_t}
\end{equation}
while the latter as;
\begin{equation}
  |\omega^{ef}_{r+\Delta r}(\tau)-\omega^{ef}_{r-\Delta r}(\tau)|/2.
  \label{eq:wef_r}
\end{equation}
Finally, both these uncertainties are added quadratically with the
existing error estimates on the scaled optical depth of each pixel to get
its final uncertainty.

\begin{figure}   
  \centering
  \includegraphics[height=7cm,width=7cm,trim={0cm 0cm 0cm
      0cm}]{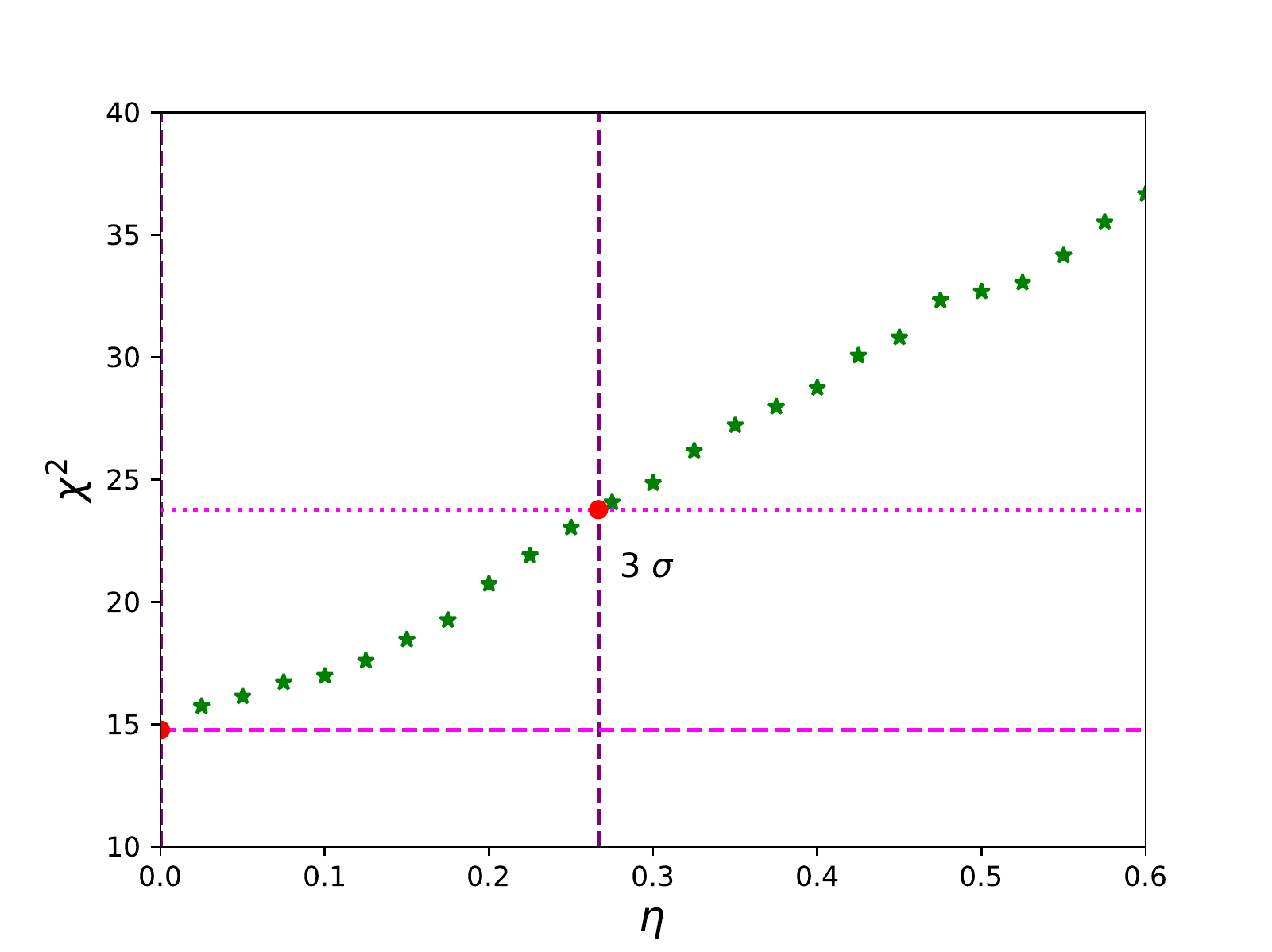}
  \caption {The plot shows the $\chi^2$ values calculated between
     $\tau_{prox}\times
    [1+\omega^{ef}_r(\tau,\omega_r)]/\tau_{\textsc{igm}}$ of LPE and
    $\tau_{prox}\times [1+\omega^{ef}_r(\tau,\eta
      \omega_r)]/\tau_{\textsc{igm}}$  of TPE, where $\eta$ is the
      fraction of the illumination in the transverse direction as
      compared to the longitudinal direction, ranging in $0 \leq \eta
      \leq 1$ (see text). The horizontal dotted line gives the
      $\chi^2$ that corresponds to 3$\sigma$ range of $\eta$ around
      the best fit value.}
 \label{Fig:k_vs_chi2}
\end{figure}

 This procedure of calculating the ionization correction is repeated
 for all pixels in the longitudinal direction. This has allowed us to
 plot the photoionization corrected excess overdensity profile along
 the longitudinal direction as shown in the right panel of
 Fig.~\ref{Fig:omega_crct_overdensity} (shown by red, diamond).
   The trend in this figure suggests the existence of an ``excess
   overdensity" at $r \leq 5$ Mpc of the quasar. We note that the
   impact of peculiar velocities might be significant in these radial
   bins which may make our distance estimation uncertain.  However,
   given the similarity of this ``excess overdensity" with that of the
   transverse direction for which the quasar’s ionization correction
   is not applied, shows that the quasar's illumination in the
   transverse direction has to be smaller in comparison to the
   illumination measured in the longitudinal direction (estimated
   based on measured luminosity in the longitudinal direction).
 Therefore, for the ionization correction in the transverse direction,
 we have followed the above procedure by replacing $\omega_r^{i}$ with
 $\eta \omega_r^{i}$, for an allowed range of $\eta$ from zero to
 unity.  For computing the best fit value of $\eta$, we have used
 $\chi^2$ analysis over 0-15 Mpc radial distance bins to
 statistically match the distribution of $\tau_{prox}\times
 [1+\omega^{ef}_r(\tau,\omega_r)]/\tau_{\textsc{igm}}$ in the
 longitudinal direction with various distributions of
 $[1+\omega^{ef}_r(\tau,\eta \omega_r)]/\tau_{\textsc{igm}}$ in the
 transverse direction by varying the trial values of $\eta$ from zero
 to unity in steps of 0.025. Here, we limit this comparison only
   to the pixels with $z_a < z_f$ (i.e., only among regions towards
   the observer). This results in a $\chi^2$ versus $\eta$ curve as
 shown in Fig.~\ref{Fig:k_vs_chi2}. As evident from this plot, the
 minimum value of $\chi^2$ is found for $\eta=0$ with
 $\chi_{min}^2=14.76$. The value of $\chi^2_{min}+n^2$ in this
 $\chi^2$ versus $\eta$ curve is used to estimate the $n\sigma$
 confidence in the $\eta$ value (assuming the Gaussian nature of the
 errors) using cubic spline interpolation. This results in $\eta
 \le 0.03, 0.16$ and $0.27$ at 1$\sigma$, 2$\sigma$ and 3$\sigma$
 level respectively. This suggests that the quasar's average
 illumination in our sample on the H~{\sc i} cloud in the transverse
 direction is less than 27\% (at 3$\sigma$ confidence level) of
 that measured along the longitudinal direction. The above
   3$\sigma$ limit changes from the 27\% value to 31\% if we exclude
   the first radial bin (i.e., 0-1 Mpc) from our above analysis, in
   view of the possible significant impact of the peculiar velocities
   close to the quasars.

\subsubsection{Impact of Ly continuum uncertainty based on spectral index variation}
  \label{s:alpha_var}
\begin{figure}   
   \centering
   \includegraphics[height=8cm,width=9cm,trim={0cm 0cm 0cm 0cm}]{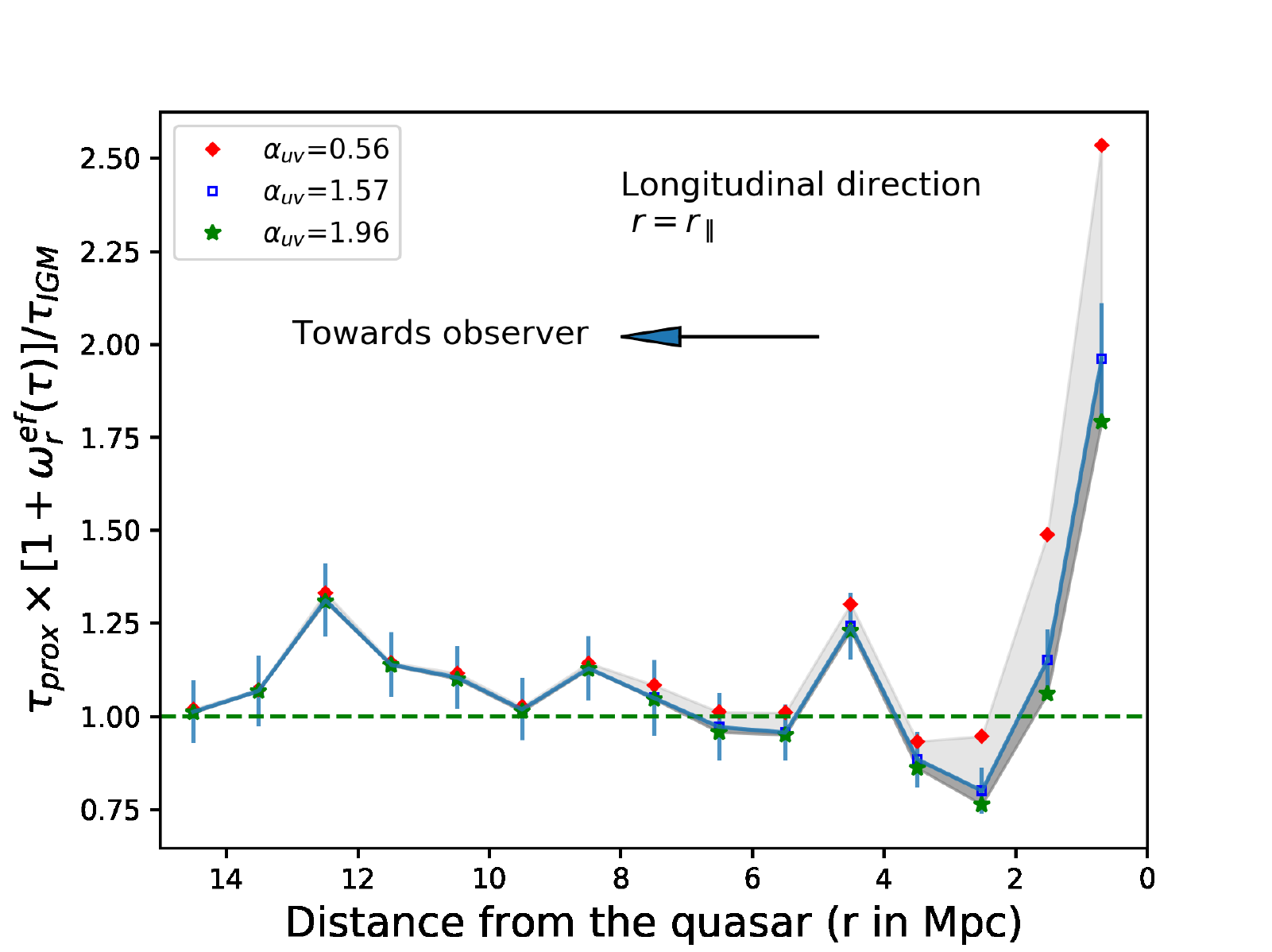} 
  \caption {The plot shows the variation of excess overdensity in the
    longitudinal direction by taking two possible extreme values of
    the spectral index while correcting for the ionization by the
    foreground quasars.}
      \label{Fig:alpha_overdensity}
\end{figure}

 As discussed in Sect.~\ref{s:distance}, we have adopted a simplistic
 treatment of the quasar SED, by using a simple power-law continuum
 composite of AGN spectrum (e.g., see Eq.~\ref{eq:power}) with no
 scatter in this relation. Any scatter around this spectral slope
 will, in fact, affect the ionization scaling, being it inversely
 proportional to the spectral slope of the quasar (e.g., see
 Eq.~\ref{eqn:gamma_q}). Therefore, in order see the effect of
 spectral slope uncertainty on the inferred overdensity, we have
 repeated our LPE analysis using two extreme values of the spectral
 slope viz., $\alpha_{uv}=$ 0.56 and 1.96, based on the range given by
 \citet[][]{Khaire2017MNRAS.471..255K}, assuming
 $\Gamma_{\textsc{uvb}}$ to be unchanged. We have re-calculated the
 overdensity profile along the longitudinal direction using these two
 extreme values, as shown in Fig.~\ref{Fig:alpha_overdensity}. As can
 be seen from this figure that the profile of overdensity is similar
 for these two extreme values of the spectral slope as is found using the
 optimal value of $\alpha_{uv}=1.57$ (except the change in
 amplitude). In addition to this, their corresponding difference in
 the amplitude of overdensity (i.e., higher for smaller $\alpha_{uv}$
 value) is also consistent within 1$\sigma$, except in lower radial
 distance bin (approaching up to 2$\sigma$) as expected due to the
 $r^{-2}\times \alpha^{-1}$ dependency of $\Gamma_q$ (e.g., see
 Eq.~\ref{eqn:gamma_q}). Moreover, we also notice that the actual
 deviation, based on the use of an individual $\alpha_{uv}$ value of
 the quasars will be smaller than that estimated here based on these
 two extreme values. As a result, we can conclude that any uncertainty
 due to the use of constant $\alpha_{uv}$ value for all our foreground quasars will not have any
 significant impact on our results. However, this will add a small
 scatter to the derived density profile.

 \section{Discussions}
 \label{s:Discussion}
Our detailed analysis of the proximity effect based on the transmitted
flux measured from closely spaced quasar pairs, detected in the SDSS,
shows a clear difference in the transmitted flux profile along the
longitudinal and the transverse directions, having more H~{\sc i}
absorption in the latter case (e.g., see
Fig.~\ref{Fig:rad_flux_red}). This is also reflected in our pixel
optical depth analysis (e.g., see Fig.~\ref{Fig:cpdf_tpe}). After
  applying appropriate ionization corrections in the longitudinal
  direction, we have derived ionization corrected average excess density
  profile which shows excess up to $r\leq 5$
  Mpc. Surprisingly, it matches with the uncorrected average
  excess overdensity profile in the transverse direction (e.g., see
  Fig.~\ref{Fig:omega_crct_overdensity}). This led to an important
  result that the H~{\sc i} absorbing clouds in the transverse
  direction on an average receive the quasar's illumination $\leq$27\%
  (at 3$\sigma$ confidence level) as compared to that along the
  longitudinal direction (e.g., see Fig.~\ref{Fig:k_vs_chi2}). Below,
  we discuss our results, in the context of excess overdensity as well
  as anisotropy in the ionizing radiation field.

      \subsection{Overdensity}
           \label{s:overdens}
 It can be noted from the right panel of the
 Fig.~\ref{Fig:omega_crct_overdensity} that our analysis shows an
 excess overdensity up to $r\leq 5$ Mpc with an increasing amplitude
 towards the foreground quasar. The evidence of overdensity inferred
 in our analysis is also found to be consistent with many previous
 such studies
 \citep[][]{DOdorico2002A&A...390...13D,Rollinde2005MNRAS.361.1015R,Guimaraes2007MNRAS.377..657G,Prochaska2013ApJ...776..136P,Adams2015MNRAS.448.1335A}.
 However, we note that the extent of the overdensity around the
 quasars found in our analysis is smaller than those found using LPE
 by \citet[][]{Rollinde2005MNRAS.361.1015R} and
 \citet[][]{Guimaraes2007MNRAS.377..657G}, though the magnitude of
 overdensity is quite similar at smaller distances.  One of the
 reasons for this difference could be that the quasars used in their
 sample are systematically brighter than the quasars used in our
 sample.  Also, these studies use quasar's spectra with better
 spectral resolution and SNR. This was affordable in these studies, as
 their main aim was only to probe the longitudinal proximity region
 for which any high-quality spectrum can be included without
 satisfying an additional requirement of being a member of closely
 separated quasar pair as needed for the TPE analysis. \par We
 constrain the overdensity profile along the longitudinal direction
 using the quasar luminosity and optimal ionization correction
 obtained using our simulations. This can in principle be used to
 place constraints on the mass of the host galaxies. However, a more
 precise constraint can also be achieved by the longitudinal proximity
 analysis based on a larger sample (affordable due to the absence of
 an additional requirement of closely separated quasar pairs) and
 realistic simulations that also models quasars along with their
 ionization influence. The coarse sampling of 1 Mpc achieved in our
 analysis based on these 181 quasars pairs, may not be enough to
 resolve the mass information which is mostly confined to $r\leq$1 Mpc
 \citep[e.g., see][]{Faucher2008ApJ...673...39F}. Additionally,
   the ionization correction in 0-1 Mpc bin might also be highly
   uncertain due to the possible impact of peculiar velocities close
   to the quasar. Therefore, in what follows, we basically focus on
 the ionization anisotropy.

 \subsection{Anisotropic distribution of radiation?}
 \label{s:anisotropy}

The difference in the transmitted flux and the optical depth found
between the longitudinal and the transverse direction, is also found
consistent with the studies of
\citet[][]{Prochaska2013ApJ...776..136P} and
\citet[][]{Kirkman2008MNRAS.391.1457K}, though based on a slightly
different analysis. A few noticeable differences are that
\citet[][]{Prochaska2013ApJ...776..136P} have used a control sample
for the study of TPE and reported the anisotropy by comparing with a
previous study of LPE which is inferred from a different quasar
sample. However, \citet[][]{Kirkman2008MNRAS.391.1457K} studied both
the LPE and TPE using the same set of the foreground quasars (similar
to our analysis), but instead of using a control sample they have
carried out the analysis correcting the observed optical depth for the
redshift evolution.  \par The inferred anisotropy in the transmitted
flux and the optical depth can have many physical interpretations,
such as, either an intrinsic anisotropy of the emitted radiation or
due to asymmetric obscuration by a ``dusty torus'' in the context of
standard AGN-unification scheme \citep[e.g.,
  see][]{Antonucci1993ARA&A..31..473A,Elvis2000ApJ...545...63E}. The
latter scenario seems to be more likely as there have been a number of
observational evidence which directly detects the presence of an
obscuring torus surrounding a central continuum source \citep[e.g.,
  see][]{Davies2015ApJ...806..127D,Almeyda2017ApJ...843....3A}.  This
will result in the obscuration of the ionizing photons in the
direction of the equatorial plane of the torus.  However, such an
interpretation will be ruled out if the quasar sample used in our
analysis have random orientations relative to our line-of-sight. \par
On the other hand, in any flux-limited survey, the quasars would be
preferentially observed along the direction in which they appear to be
brighter such as in the direction perpendicular to the plane of the
dusty torus. This could be a likely scenario, as for the majority of
SDSS DR12 quasars the redshift measurements are derived from ``broad
emission lines'' \citep[e.g., see
][]{Alexandroff2013MNRAS.435.3306A,Paris2017A&A...597A..79P} which are
prominent in the quasars having their torus perpendicular to our
line-of-sight, commonly known as Type-I AGNs (in the standard
AGN-unification scheme).  Therefore, we checked the velocity
  width (FWHM) of the C~{\sc iv} line as provided by SDSS \citep[e.g.,
    see ][]{Paris2017A&A...597A..79P}. Out of 181 foreground quasars
  in our sample, it was available for 175 of them (6 of them did not
  have prominent C~{\sc iv} emission line) and indeed all of them have
  FWHM $>2000$ \kms. This suggests that our sample is dominated by
  Type-I AGNs.  It will also be interesting to extend our analysis
  (based on Type-I AGNs as foreground quasars) by using the Type-II
  AGNs as foreground quasars, in-spite of the scarcity of observed
  Type-II AGNs at higher redshifts \citep[see
    also,][]{Lu2011ApJ...736...49L}. This will confirm the scenario of
  anisotropic obscuration, if one finds an exactly opposite trend as
  compared to our above analysis, due to the presence of dusty torus
  along the longitudinal direction in Type-II AGNs.

 An alternative possibility for the observed anisotropy among the
 longitudinal and the transverse directions could be the finite
 lifetime of the quasar. As discussed in Sect.~\ref{s:Intro} that the
 light from the foreground quasars has to travel an extra path in the
 transverse direction (i.e., $r_\perp$/c). Therefore, due to quasar's
 finite lifetime, its radiation might not be apparent along the
 background quasar when we see the foreground quasar in its initial
 stages of the active phase. We use the median $r_\perp$/c of our
 sample i.e., 0.5 Mpc, which results in an extra light travel time of
 0.5 Mpc/c $\sim$ 1.6 Myr. In addition to this, the fact that we
 observe the overdensity even in the first radial bin of the
 transverse direction implies that the quasar was not ionizing before
 1.6 Myr ago i.e., quasar lifetime $t_Q \leq 1.6$ Myr. \par
   Moreover, the estimate of quasar's lifetime reported in literature,
   based on various techniques, varies over a wide range from 10$^4$
   yrs to 10$^7$ yrs \citep[e.g., see
   ][]{Srianand_1997,Goncalves2008ApJ...676..816G,Trainor2013ApJ...775L...3T,
     Borisova_2016,Oppenheimer10.1093/mnras/stx2967}. For instance,
   \citet[][]{Bajtlik1988ApJ...327..570B} estimated the quasar's
   lifetime to be around 10$^4$ yrs using the equilibration
   time-scale $\sim \Gamma_{\textsc{H~{\sc i}}}^{-1}$ \citep[see
     also,][]{Carswell1982MNRAS.198...91C,Agliolpe32008AA...480..359D,Eilers_2017,Eilers_2018}.
   However, based on the quasar variability,
   \citet[][]{Schawinski2015MNRAS.451.2517S} estimated the typical
   value to be 0.1 Myr \citep[see
     also,][]{Schawinski_2010ApJ.724.1S}. On the other extreme,
   \citet[][]{Schmidt2018ApJ...861..122S} have used longitudinal
   He~{\sc ii} proximity effect and estimated quasar's lifetime to be
   $\sim$30 Myr \citep[see
     also,][]{Hogan1997AJ....113.1495H,Anderson_1999,Zheng_2015,Khrykin2016ApJ...824..133K,Schmidt_2017}.
   Given the wide range of inferred values of quasar's lifetime it is
   difficult to draw any conclusion on the value of 1.6 Myr as we have
   estimated above. \par Another possible scenario could be the
 inflow of matter onto the quasar massive halo \citep[see
   also,][]{Faucher2008ApJ...673...39F}. This could also introduce
 anisotropic effects on the measured redshift of the absorbing
 clouds. For instance, due to the inflow, the redshift distortion will
 be more in the longitudinal direction as compared to the transverse
 direction, with the latter being affected by only the projected
 velocity component of the inflow. As pointed out by
 \citet[][]{Faucher2008ApJ...673...39F}, it will amount to a typical
 underestimation of the distance by 1 Mpc in the longitudinal
 direction which is too small to explain the significant difference we
 noticed among these two directions over a range up to 4
 Mpc. Moreover, this implies that we might have applied an ionization
 correction a bit more than the actual value in the longitudinal
 direction (due to the distance underestimation) to scale up the
 observed optical depth (compared to the true ionization
 correction). However, in view of the fact that the scaled optical
 depth in the longitudinal direction does not exceed even the value of
 unscaled optical depth (i.e., without ionization correction) in the
 transverse direction, we conclude that any possible impact of an
 inflow will be negligible to the measured anisotropy. Similar
   to inflows, quasars also do have outflows, which will have an
   impact on redshift distortion opposite to that of the inflow
   scenario. However, given the large difference between the
   longitudinal and transverse directions, it seems that simply the
   effect of inflow and/or outflow cannot explain the above observed
   discrepancy inferred in the quasar's illumination among these two
   directions. \par The most probable scenario for the observed
 anisotropy (among the longitudinal and transverse directions) seems
 to be anisotropic obscuration of the ionizing radiation, probably due
 to the presence of the dusty torus. This also has an important
 implication for the AGN unification scheme. For instance,
 \citet[][]{Schirber2004ApJ...610..105S} used the absence of TPE to
 constrain such dusty torus to have an opening angle of
 $\sim$20$\degree$, and \citet[][]{Kirkman2008MNRAS.391.1457K} over a
 range of 60$\degree$ to 120$\degree$ \citep[also confirmed
   by][]{Lu2011ApJ...736...49L}. We have constrained the average
 quasar's illumination ``in the framework of the shadowing by the
 torus'' on the H~{\sc i} cloud in the transverse direction in
 comparison to the longitudinal direction to be less than 27\%
 at 3$\sigma$ confidence (see Sect.~\ref{s:over} for details). \par
 Furthermore, we also note that a comparison of the transverse
 proximity effect, towards the observer and towards the background
 quasars can allow us to probe the temporal evolution of the quasar's
 ionization (assuming on an average the matter distribution is
 spherically symmetric). For instance,
 \citet[][]{Kirkman2008MNRAS.391.1457K} found an asymmetry with more
 absorption toward background quasars and use it to constrain the
 episodic lifetime of the quasars to be $\sim$ 1 Myr. In our study, we
 do not find such a difference as can be seen from
 Fig.~\ref{Fig:rad_flux_red}. Here, we may also recall that in our
 analysis, the emission redshifts are corrected for the systematic
 redshift by 229 \kms. The shift of the peak optical depth position
 with respect to the systematic redshift of the quasar is close to the
 average systematic redshift uncertainty provided by
 \citet[][]{Shen2016ApJ...831....7S} for the SDSS quasars.  This once
 again reiterate the importance of accurately measuring the systematic
 redshift of the quasars. The absence of such systematic correction in
 the emission redshift would have introduced any fiducial asymmetry
 (e.g., see Fig.~\ref{Fig:rad_flux}) and hence may lead to an
 erroneous interpretation in constraining the episodic quasar
 lifetime.

\section{Summary and Conclusions}
\label{s:summary}
The proximity effect is a very powerful tool to probe the environment
of the quasars as well as to measure the intensity of the diffuse UVB
radiation. Many of the past studies for the proximity effect have
found significantly higher H~{\sc i} absorption near the quasar as
compared to the expected value based on the statistics of faraway IGM
\citep[][]{Croft2004ApJ...610..642C,Schirber2004ApJ...610..105S,Rollinde2005MNRAS.361.1015R,Guimaraes2007MNRAS.377..657G,Kirkman2008MNRAS.391.1457K,Prochaska2013ApJ...776..136P,Finley2014A&A...572A..31F,Adams2015MNRAS.448.1335A}. In
addition, \citet[][]{Prochaska2013ApJ...776..136P} and
\citet[][]{Kirkman2008MNRAS.391.1457K} have also reported discrepancy
in the H~{\sc i} absorption along the longitudinal and the transverse
directions in their sample of quasar pairs. The absorption was found
to be stronger in the transverse direction. To probe the environment
of the quasars at even smaller scales, both in the longitudinal and
the transverse directions, we have used a sample of 181 projected
quasar pairs from SDSS DR12 having a small angular separation of
$<1.5$~\ar~(e.g., see Fig.~\ref{Fig:qso_prop}).  \par Our sample of
quasar pairs is compiled using proper selection criteria on the
redshift range and angular sky separation. We also removed special
cases such as: (i) BAL quasars, (ii) sightlines which have associated
absorbers, LLS, DLAs or sub-DLAs in the proximity region (e.g., see
Sect.~\ref{s:sample}) and (iii) pairs having velocity separation $<
2000$ \kms~etcetera (e.g., see Sect.~\ref{s:zq}). \par This well
selected quasar sample is analyzed with noticeable improvements such
as: \ben
\item In order to compare the distribution of the optical depth and
  the transmitted flux measured in the proximity region both along the
  longitudinal and the transverse directions, we have used the same
  set of foreground quasars.
\item For comparison with the IGM, we have used the redshift and SNR
  matched control sample, instead of using the redshift scaling to
  take into account the redshift evolutions of the IGM optical depth
  as used in many previous such studies (e.g.,
  Sect.~\ref{s:control_sample}, Fig.~\ref{Fig:z_snr_match}).
\item We have considered all the possible sources of errors in the
  flux and optical depth analysis such as (i) error contributed by
  the continuum fitting uncertainty using mock spectra for each sightline in
  our sample (e.g., see Sect.~\ref{s:conti} and
  Figs.~\ref{Fig:conti_max},~\ref{Fig:conti_min},~\ref{Fig:delc_z_snr}),
  (ii) the error due to photon counting statistics, (iii)
  sightline-to-sightline variance, (iv) emission redshift measurement
  error and (v) the r.m.s statistical error within the 1 \Mpc~radial
  distance bin (e.g., Sect.~\ref{s:Uncertainties}).
\item The observed optical depth in the proximity region can be
  enhanced due to the excess overdensity in which the quasar resides
  as well as be reduced by the excess quasar ionization. To lift this
  degeneracy between the quasar's ionization and overdensity, we use
  the precise UVB radiation measurements from
  \citet[][]{Khaire2015MNRAS.451L..30K} in conjunction with the known
  quasar luminosity in the longitudinal direction. However, the
  analytical [$1+\omega_r$] ionization correction (i.e.,
  Eq.~\ref{eqn:omega}-\ref{eqn:off}) for low/moderate resolution and
  low SNR is not found to be appropriate (e.g., see
  Sect.~\ref{s:ionization}). Therefore, we have used detailed
  hydrodynamical simulation to validate ionization correction (i.e.,
  $[1+\omega^{ef}_r(\tau,\omega_r)]$) optimal for low/moderate
  resolution spectra (e.g., see Sect.~\ref{s:simulation} and
  Figs.~\ref{Fig:simu_sightline},~\ref{Fig:w_fw}).  \een

  \par Our detailed
  analysis of the transmitted flux and the pixel optical depth, both
  in the longitudinal and the transverse directions, led us to the
  following main conclusions: \ben
  \item In the longitudinal direction, the proximity region within 4
    \Mpc~(proper distance) shows an enhancement of the transmitted
    flux in comparison to its control sample (e.g., see
    Sect.~\ref{s:Analysis1}, Fig.~\ref{Fig:rad_flux_red}), while the trend
    is found to be reversed in the transverse direction ($r \leq 2$
    Mpc). This suggests the dominance of the
    quasars ionization in the longitudinal direction whereas higher
    H~{\sc i} absorption in the transverse direction. A corresponding
    consistent trend is also seen in the CPDFs of the effective
    optical depth in different radial distance bins from the
    foreground quasar (e.g., see Sect.~\ref{s:op_dep},
    Fig.~\ref{Fig:cpdf_tpe}).
\item The average absorption profile we measured in the
  transverse direction is found to be symmetric with respect to the
  quasar systematic redshift (e.g., see
  Figs.~\ref{Fig:rad_flux_red},~\ref{Fig:rad_flux}). We also show the
  actual profile is sensitive to the assumed quasar
  redshifts. Therefore, to make an important progress in mapping the
  density profile, NIR spectra based systematic redshift measurements
  are imperative.
    
\item We estimated the quasar's ionization based on the measured
  luminosity in the longitudinal direction to get the ionization
  corrected pixel optical depth distribution. This enabled us to
  estimate the average excess overdensity profile (after ionization
  correction), which is found to be decreasing radially outward from
  the quasars and with an excess up to $r \leq 5$ Mpc.

\item Unlike the longitudinal direction, due to the lack of direct
  measurement of the quasar's luminosity in the transverse direction,
  the direct estimation of its average excess overdensity profile
  (after ionization correction) was not feasible. However, it is
  reasonable to assume the overdensity distribution to be symmetric as
  we average over a large number of quasars. This in conjunction with
  the fact that we observed more H~{\sc i} absorption in the
  transverse direction clearly suggests that the quasar's obscuration
  in our sample is more in the transverse direction.  This has allowed
  us to constrain the quasar's illumination on H~{\sc i} absorbing
  cloud in the transverse direction to be $\leq$27\% (3$\sigma$
    confidence level) of that along the longitudinal direction (e.g.,
  see Fig.~\ref{Fig:k_vs_chi2}). The above scenario of anisotropic
  obscuration after excluding many other possibilities (e.g., see
  Sect.~\ref{s:anisotropy}) is also found consistent with the fact
  that all of our foreground quasars happen to be Type-I AGNs (based
  on the velocity width of C~{\sc iv} emission line) for which dusty
  torus will be mainly in the transverse direction. This also
  independently confirms the presence of a dusty torus in the AGNs,
  which is a key feature of the AGN unification scheme.

 \een
  
\par For further improvement, it will also be interesting to extend
our analysis (based on Type-I AGN as foreground quasars) by using the
Type-II AGN as well, beside enlarging the sample size of such quasar
pairs at separation $< 1$ \ar~\citep[as also suggested
  by][]{Lu2011ApJ...736...49L} The improvement in the sample size will
require a dedicated program of long-slit spectroscopy with 3-4 m class
telescopes by orienting the slit to cover both members of a probable
quasar pair even at separation $<$ 1 \ar~which might have been missed
by SDSS observations due to fibre collision problem; so as to probe
the quasar lifetime, anisotropic obscuration and their environment
even at \kpc~scales.
 
\section*{Acknowledgements}
 We thank the anonymous referee for his/her valuable comments and
 constructive suggestions which has greatly improved the manuscript.
 We gratefully acknowledge Vikram Khaire for providing us with the
 latest value of UV-background radiation and Prakash Gaikwad for
 providing the simulated IGM spectra. We are also thankful to
 Tirthankar Roy Choudhury for useful discussions.  \par Funding for
 the SDSS and SDSS-II has been provided by the Alfred P.  Sloan
 Foundation, the Participating Institutions, the National Science
 Foundation, the U.S. Department of Energy, the National Aeronautics
 and Space Administration, the Japanese Monbukagakusho, the Max Planck
 Society, and the Higher Education Funding Council for England. The
 SDSS Web Site is http://www.sdss.org/. The SDSS is managed by the
 Astrophysical Research Consortium for the Participating
 Institutions. The Participating Institutions are the American Museum
 of Natural History, Astrophysical Institute Potsdam, University of
 Basel, University of Cambridge, Case Western Reserve University,
 University of Chicago, Drexel University, Fermilab, the Institute for
 Advanced Study, the Japan Participation Group, Johns Hopkins
 University, the Joint Institute for Nuclear Astrophysics, the Kavli
 Institute for Particle Astrophysics and Cosmology, the Korean
 Scientist Group, the Chinese Academy of Sciences (LAMOST), Los Alamos
 National Laboratory, the Max-Planck-Institute for Astronomy (MPIA),
 the Max-Planck-Institute for Astrophysics (MPA), New Mexico State
 University, Ohio State University, University of Pittsburgh,
 University of Portsmouth, Princeton University, the United States
 Naval Observatory, and the University of Washington.

 \bibliography{references}
\end{document}